\documentclass[prc,letterpaper,twocolumn,showpacs,showkeys,lengthcheck,
  nofootinbib,tightenlines,preprintnumbers,superscriptaddress,
  ]{revtex4-1}

\usepackage[utf8]{inputenc}
\usepackage{newtxtext,newtxmath}

\usepackage{graphicx}
\usepackage[usenames,dvipsnames]{xcolor}

\usepackage{hyperref}
\hypersetup{colorlinks=true,citecolor=blue,linkcolor=blue,urlcolor=blue}

\usepackage{subcaption}
\usepackage{amsmath,amssymb,amsfonts}

\usepackage{slashed}

\begin{document}
\title{
  Extraction of $t$-slopes from experimental
  $\gamma p\rightarrow K^+\Lambda$ and
  $\gamma p\rightarrow K^+\Sigma^0$
  cross section data
}
\author{Adam Freese}
\affiliation{Argonne National Laboratory, Lemont, IL 60439}
\affiliation{Florida International University, Miami, FL 33199}
\author{Daniel Puentes}
\author{Shankar Adhikari}
\author{Rafael Badui}
\author{Lei Guo}
\author{Brian Raue}
\affiliation{Florida International University, Miami, FL 33199}
\date{\today}

\begin{abstract}
  We analyze recent $K^+$ meson photoproduction data from
  the CLAS collaboration
  for the reactions
  $\gamma p\rightarrow K^+\Lambda$ and
  $\gamma p\rightarrow K^+\Sigma^0$,
  fitting measured
  forward-angle differential cross sections
  to the form
  $Ae^{Bt}$.
  We develop a quantitative scheme for determining the kinematic region
  where the fit is to be done,
  and, from the extracted $t$-slope $B$,
  determine whether single-Reggeon exchange can explain the
  production mechanism.
  We find that,
  in the region $5 < s < 8.1$~GeV$^2$,
  production of the $K^+\Lambda$ channel can be explained by
  single $K^+$ Reggeon exchange,
  but the $K^+\Sigma^0$ production channel cannot.
  We verify these conclusions by fitting the data to
  a differential cross section produced by the
  interfering sum of two exponential amplitudes.
\end{abstract}

\maketitle



\section{Introduction}
\label{sec:intro}

In recent years,
the CEBAF Large Acceptance Spectrometer (CLAS) collaboration
at Jefferson Lab
has collected a large volume of high-precision data for
the unpolarized photoproduction of $K^+$ mesons from a proton target
with a $\Lambda$ or $\Sigma^0$ hyperon as the
recoil baryon~\cite{Bradford:2005pt,McCracken:2009ra,Dey:2010hh}.
The experiments
measured the differential cross section $\frac{d\sigma}{dt}$ for
a range of low and intermediate
photon energies from $E_\gamma=0.91$~GeV to $E_\gamma=3.83$~GeV,
corresponding to squared center-of-momentum energies
from $s=2.59$~GeV$^2$ to $s=8.07$~GeV$^2$,
and a wide range of angles from the forward to backward region.

Such a wide kinematic coverage allows for the data
to be used in testing a variety of models with different
domains of applicability.
For instance, at central angles---specifically,
$90^\circ$ in the center-of-momentum frame---the quark counting rule
can be tested~\cite{Schumacher:2010qx}.
By contrast, at forward angles, one can investigate the applicability
of diffractive scattering and production models,
such as those formulated on the basis of Regge theory.
Additionally, the $K^+$ photoproduction data has been vital
in the extraction of intermediate $s$-channel resonances
in both partial wave analyses and effective field theories
({\sl cf.}~\cite{Anisovich:2007bq,Nikonov:2007br,Anisovich:2010an,
  Anisovich:2011ye,Anisovich:2011su,Anisovich:2014yza}).
The range of photon energies at which these measurements were done
allows the tests of said models to be extended
into center-of-momentum energies below which
they have so far been successfully applied.

In this work, we look specifically at the recent CLAS data in
the diffractive, forward production regime.
Within this regime, it has been customary
({\sl cf.}\ {\sl e.g.},~\cite{Bauer:1977iq})
to fit data to an exponential function:
\begin{equation}
  \frac{d\sigma}{dt} = A(s)e^{B(s)t}
  \label{eqn:AeBt}
  .
\end{equation}
It is specifically the $t$-slope factor $B$ that we will extract.
Moreover, we will develop a quantitative scheme
for determining the range of $t$ over which
the fit to Eq.~(\ref{eqn:AeBt}) should be done.

The paper is organized as follows.
In Section \ref{sec:regge}, we briefly review some basic results
from Regge theory that justify the use of Eq.~(\ref{eqn:AeBt})
and help determine its range of validity.
In Section \ref{sec:exp}, we perform the exponential fit to recent
CLAS data for the reactions
$\gamma p\rightarrow K^+\Lambda$ and
$\gamma p\rightarrow K^+\Sigma^0$
and develop a scheme for determining
the appropriate range of $t$ values to perform the fit.
In Section \ref{sec:double}, we perform an additional fit
to the interfering sum of two exponentials
in order to further investigate the results of the prior section.
Finally, in \ref{sec:end}, we reiterate our conclusions
and consider implications of this investigation.


\section{$t$-slopes and Regge trajectories}
\label{sec:regge}

Regge theory is a phenomenological theory that explains
hadronic scattering amplitudes
using mathematical properties of the scattering matrix
in place of a fundamental theory of the strong nuclear interaction.
It relies on imposing a handful of simple properties,
namely unitarity of the $S$-matrix,
analyticity in terms of physical observables,
and crossing symmetry
in order to constrain the functional form that the scattering amplitude.

One of the most fruitful methods employed by Regge theory is to
analytically continue functions of the angular momentum quantum number $l$
into the complex plane.
This allows a partial-wave expansion of the scattering amplitude
to be rewritten as a sum of
integrals around cuts and poles in the complex-$l$ plane,
in what is known as the Sommerfeld-Watson representation.
Cuts tend to dominate the scattering amplitude at high $-t$,
while poles dominate at smaller $-t$~\cite{Collins:1977jy}.

Using the fit form in Eq.~(\ref{eqn:AeBt}) for a given $s$ value
finds justification within Regge theory,
provided that the reaction has a large center-of-momentum energy $\sqrt{s}$
and a small invariant momentum transfer $-t$.
More precisely, the conditions
$s\rightarrow\infty$ and $-t \ll s$ should hold.
The small $-t$ leads us to expect poles to dominate
the Sommerfeld-Watson representation of the scattering amplitude,
and the conditions $s\rightarrow\infty$ and $-t \ll s$ together mean
the contribution of one pole to the scattering
amplitude $\mathcal{A}(s,t)$ follows the asymptotic form~\cite{Collins:1977jy}:
\begin{equation}
  \mathcal{A}(s,t) \approx A(s)\left(\frac{s}{s_0}\right)^{\alpha_R(t)}
  \label{eqn:Ast}
  ,
\end{equation}
where $A(s)$ is a function of $s$ alone.
Here, $s_0$ is not actually constant, but a function of $t$.
However, the $t$-dependence of $s_0$ is conventionally neglected
and a central value of
$s_0=1$~GeV$^2$ is typically used~\cite{Collins:1977jy,Guidal:1997hy}.
The function $\alpha_R(t)$ is the real part of
the location of the pole in the complex-$l$ plane,
which represents infinitely many exchanged particles
with different $l$ but otherwise identical quantum numbers.

Due to crossing symmetry,
the function $\alpha_R$ should be defined for
both positive and negative values of its argument.
In particular, where $\alpha_R(t)$ appears in a $t$-channel diagram
for the process $AB\rightarrow CD$,
the expression $\alpha_R(s)$ will appear in an expression for
the ``crossed'' process $A\bar{C}\rightarrow B\bar{D}$.
When $s$ is the squared mass of a resonance with
the quantum numbers of the Regge pole in question
and internal angular momentum $j$,
{\sl i.e.}, when $s=m_j^2$, one should have $\alpha_R(m_j^2)=j$.
This allows for phenomenological extraction of
the functional form of $\alpha_R(t)$ in the $t > 0$ region,
and one typically finds $\alpha_R(t)$ to be approximately linear in this region.
The line along which these exchange particles fall,
and likewise the function $\alpha_R(t)$,
is called a Regge trajectory.
The particles falling on a Regge trajectory are formally treated as
a single fictitious particle with complex angular momentum $\alpha(t)$
called a Reggeon.

\begin{figure}
  \centering
  \includegraphics[width=.48\textwidth]{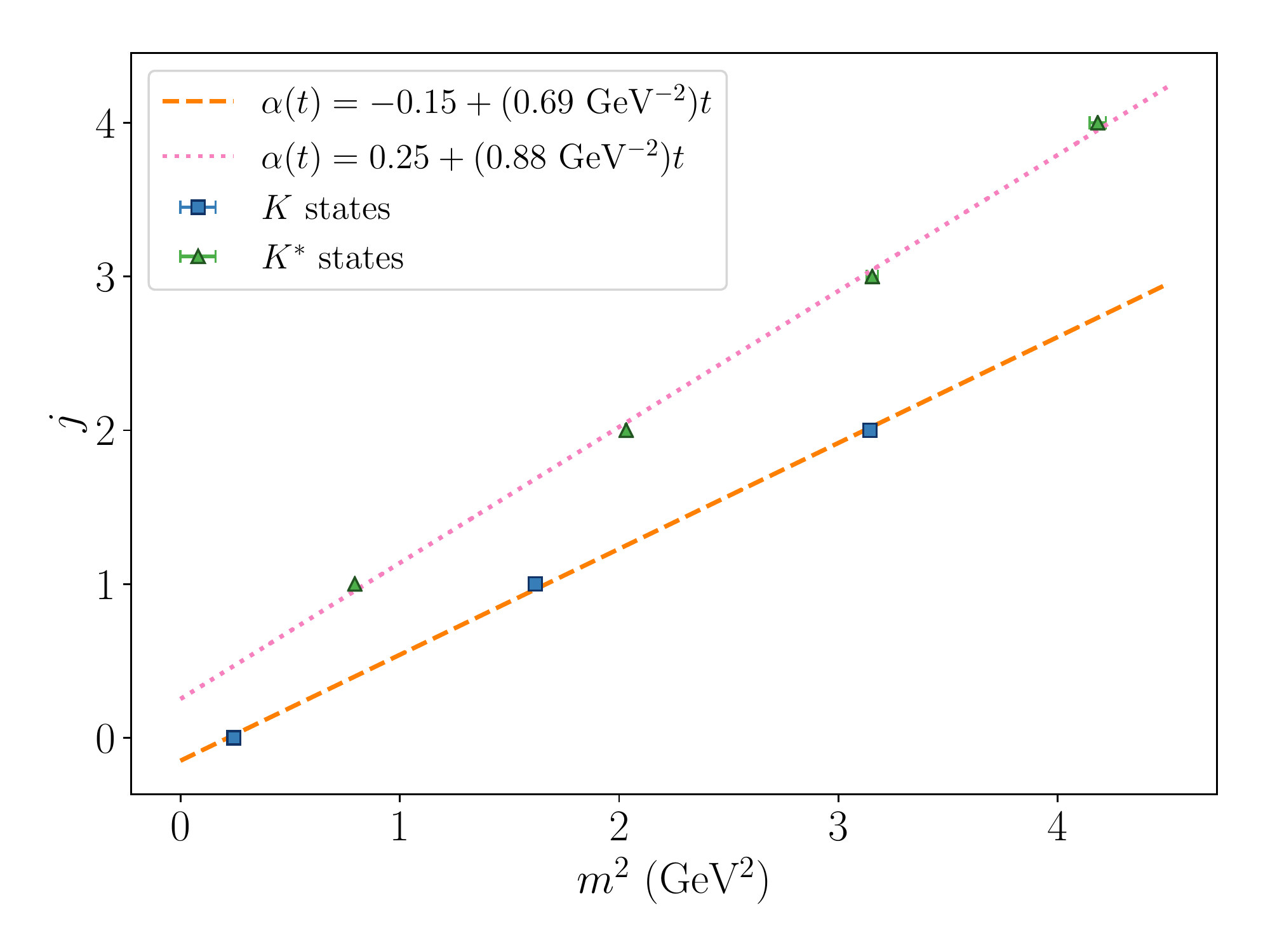}
  \caption{Regge trajectories for the $K$ and $K^*$ mesons.}
  \label{fig:regge}
\end{figure}

Relevant to the present work,
the $K$ and $K^*$ mesons fall along linear trajectories
on a $j$ vs.\ $m^2$ plot,
as can be seen in Fig.~\ref{fig:regge}.
Since analyticity is also imposed on $\alpha_R$,
the linear form
$\alpha_R(t) \approx \alpha_R(0) + \alpha^\prime_R t$
is expected to hold at least for small $-t$ as well.
It is commonly understood
(see {\sl e.g.},~\cite{Collins:1977jy,Guidal:1997hy})
that the Regge trajectory saturates for larger values of $-t$
and no longer follows this linear trend,
but this nonlinear behavior can be neglected by keeping $-t$ small,
as is needed for the asymptotic form of Eq.~(\ref{eqn:Ast})
(and the negligibility of complex-$l$ cuts)
to hold.

Using the linear form of the Regge trajectory,
along with Eq.~(\ref{eqn:Ast}),
gives the exponential fit form $Ae^{Bt}$
if we assume a single Regge trajectory
({\sl i.e.}, one pole in the complex-$l$ plane)
contributes to the overall scattering amplitude.
One moreover has
\begin{equation}
  B = 2 \alpha^\prime_R \log\left(\frac{s}{s_0}\right)
  \label{eqn:B}
  ,
\end{equation}
giving a theoretical expectation as to how
the $t$-slope we extract from the data should vary with $s$.

In the following section we will extract the $t$-slopes from
recent CLAS and older world data for
$\gamma p\rightarrow K^+\Lambda$ and $\gamma p\rightarrow K^+\Sigma^0$
(hereafter the $K^+\Lambda$ and $K^+\Sigma^0$ channels, respectively)
by fitting these data to Eq.~(\ref{eqn:AeBt}).
In light of Eq.~(\ref{eqn:B}) from Regge theory,
we will study the dependence of $B$ on the squared center-of-momentum energy, $s$,
to see if this dependence is in fact logarithmic,
and to see if the slope of $B$ versus $\log(s)$ corresponds to
either $2\alpha_K^\prime$ or $2\alpha_{K^*}^\prime$.
To find as much for either channel would suggest that
photoproduction of this channel is dominated by exchange of a single Reggeon
in the diffractive region.


\section{Exponential Fit and Slope Factors}
\label{sec:exp}

\begin{figure}
  \centering
  \begin{subfigure}[b]{.48\textwidth}
    \centering
    \includegraphics[width=\textwidth]{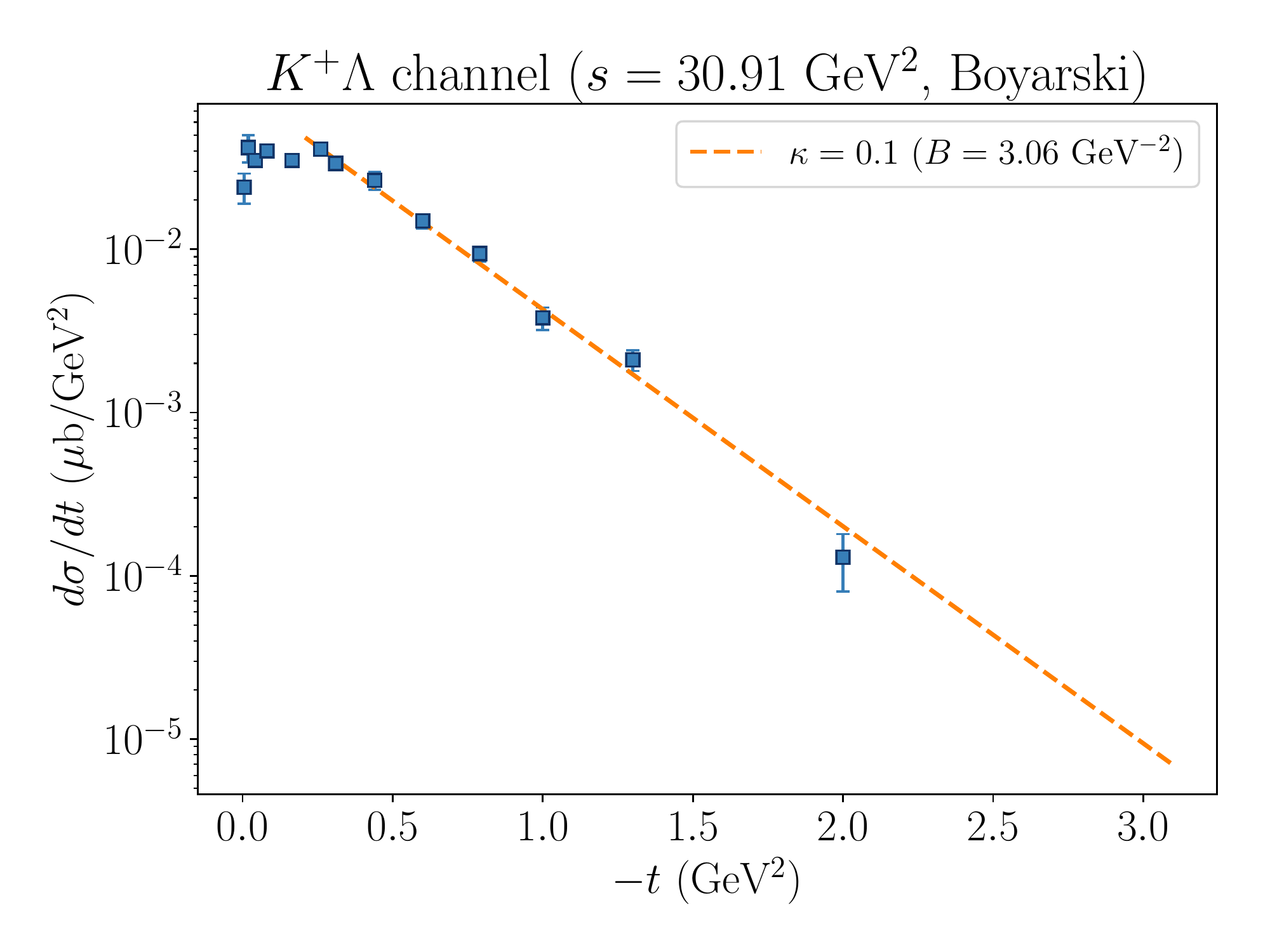}
  \end{subfigure}
  \begin{subfigure}[b]{.48\textwidth}
    \centering
    \includegraphics[width=\textwidth]{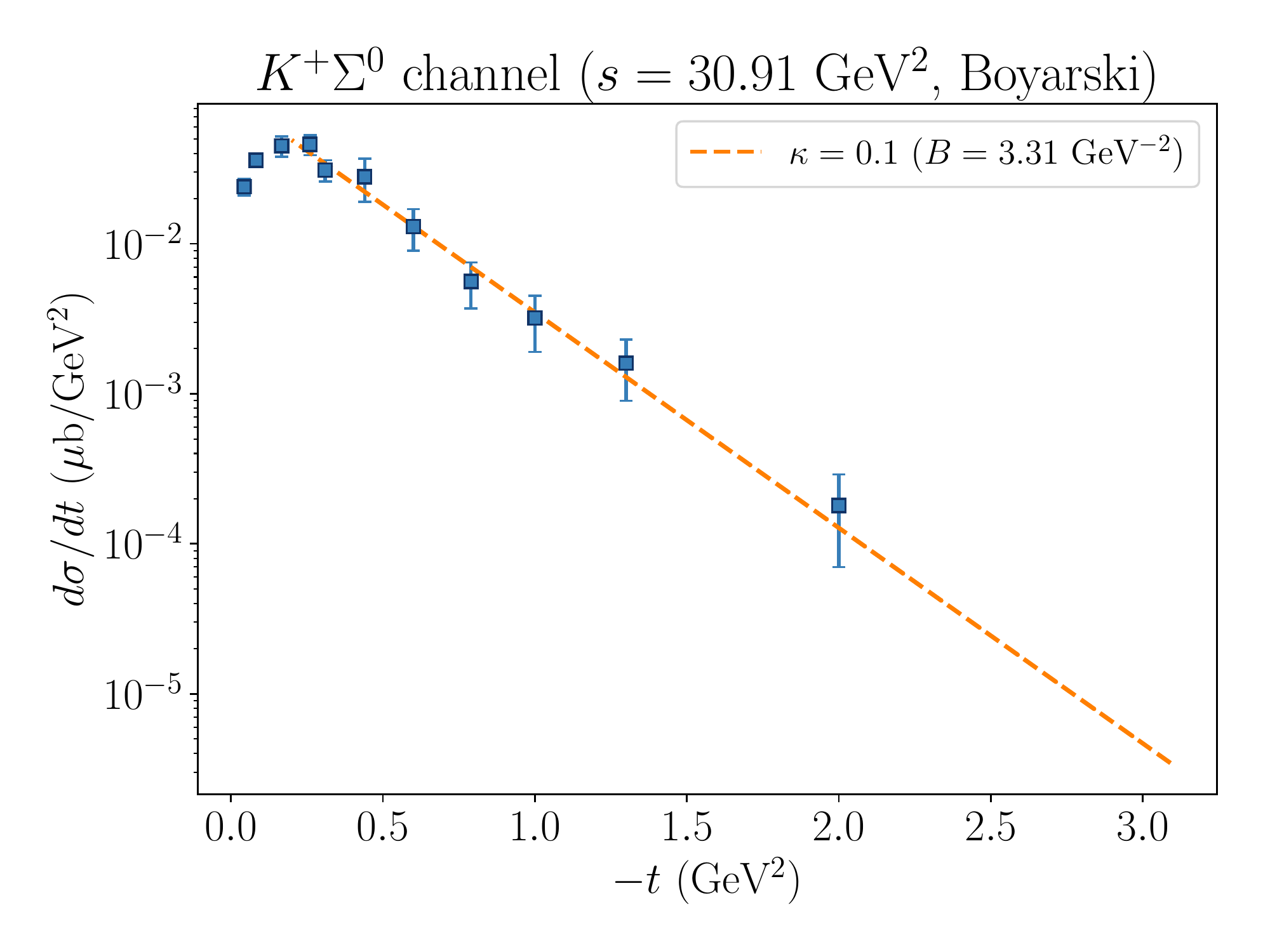}
  \end{subfigure}
  \caption{
    Differential cross section $\frac{d\sigma}{dt}$ versus $-t$ for
    (top) the $K^+\Lambda$ channel
    and
    (bottom) the $K^+\Sigma^0$ channel
    for a high squared center-of-momentum energy $s=30.91$~GeV$^2$,
    together with a fit to Eq.~(\ref{eqn:AeBt}).
    SLAC data are given in the plots~\cite{Boyarski:1970yc}.
  }
  \label{fig:highs}
\end{figure}

As explained in the previous section,
Eq.~(\ref{eqn:AeBt}) is valid for a fixed, large value of $s$
under the condition $-t \ll s$.
However, there is currently no fixed quantitative scheme for determining
what range of $t$  the condition $-t \ll s$ corresponds to.
Typically, at sufficiently high $s$,
one can determine the range for which this fit is valid
by visual inspection of a log-scaled plot.
For instance, in Fig.~\ref{fig:highs},
one can see an unambiguous straight line when
$\frac{d\sigma}{dt}$ is plotted against $-t$
with the $y$-axis logarithmically scaled.

\begin{figure}
  \centering
  \begin{subfigure}[b]{.48\textwidth}
    \centering
    \includegraphics[width=\textwidth]{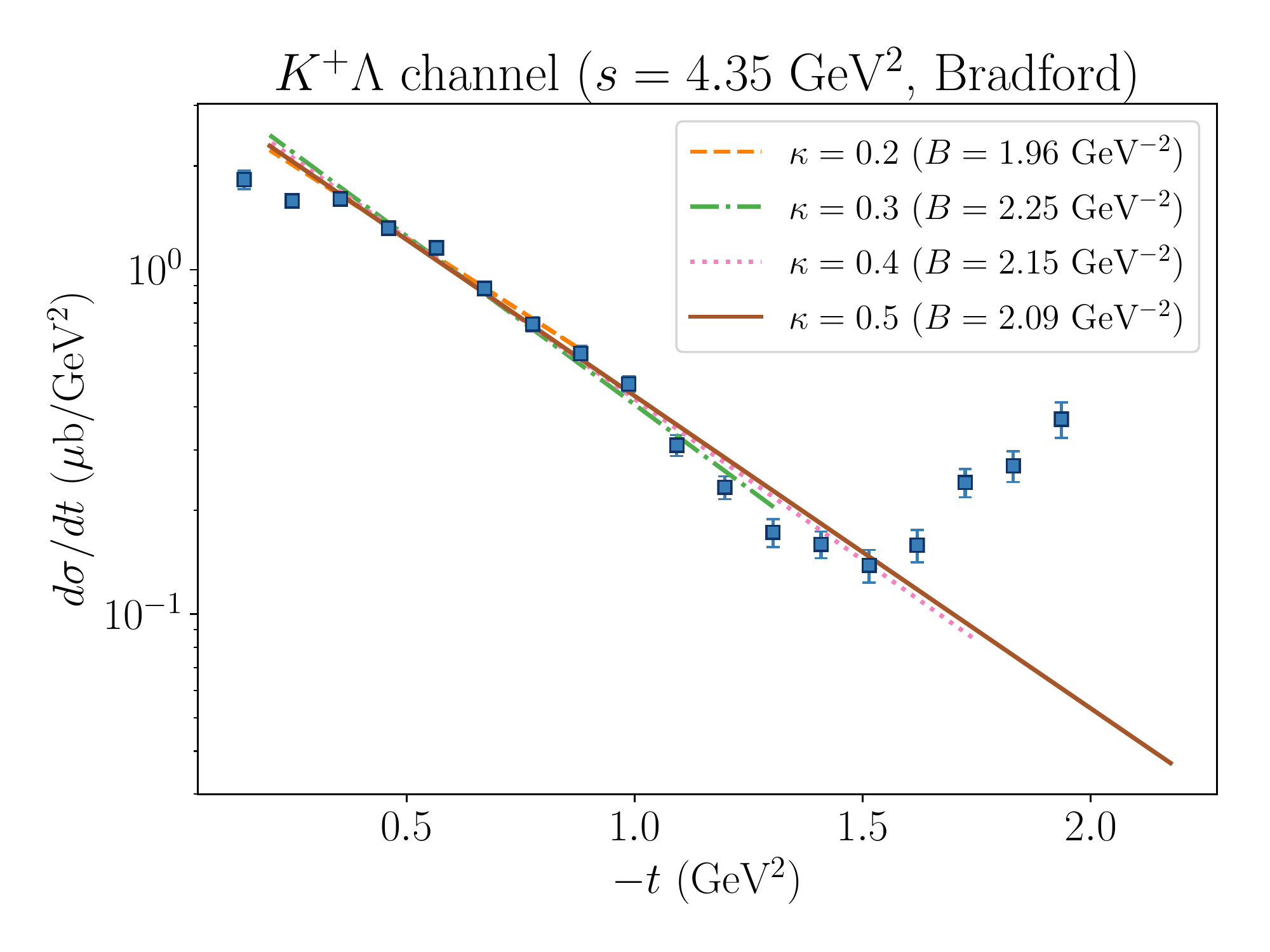}
  \end{subfigure}
  \begin{subfigure}[b]{.48\textwidth}
    \centering
    \includegraphics[width=\textwidth]{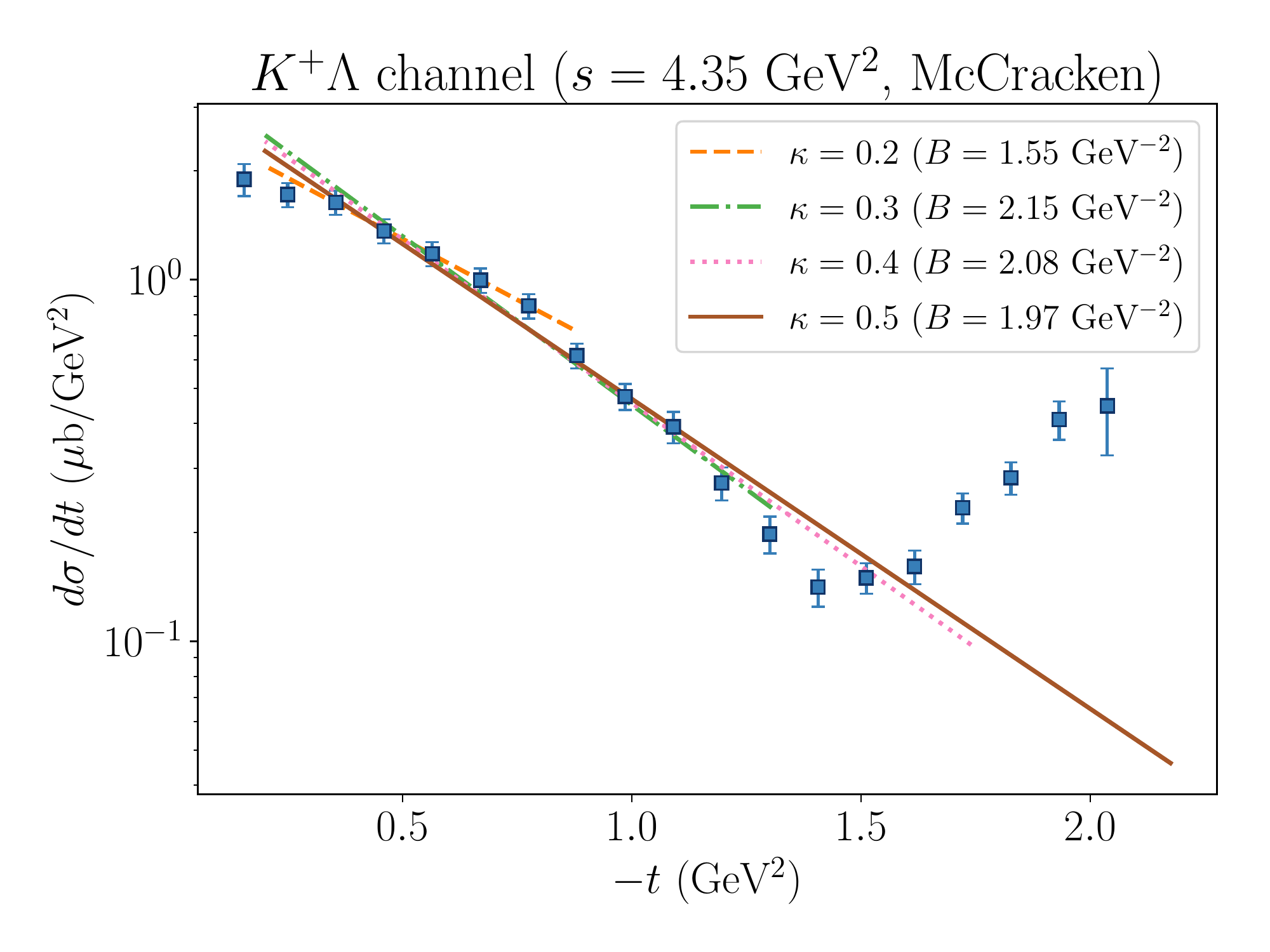}
  \end{subfigure}
  \begin{subfigure}[b]{.48\textwidth}
    \centering
    \includegraphics[width=\textwidth]{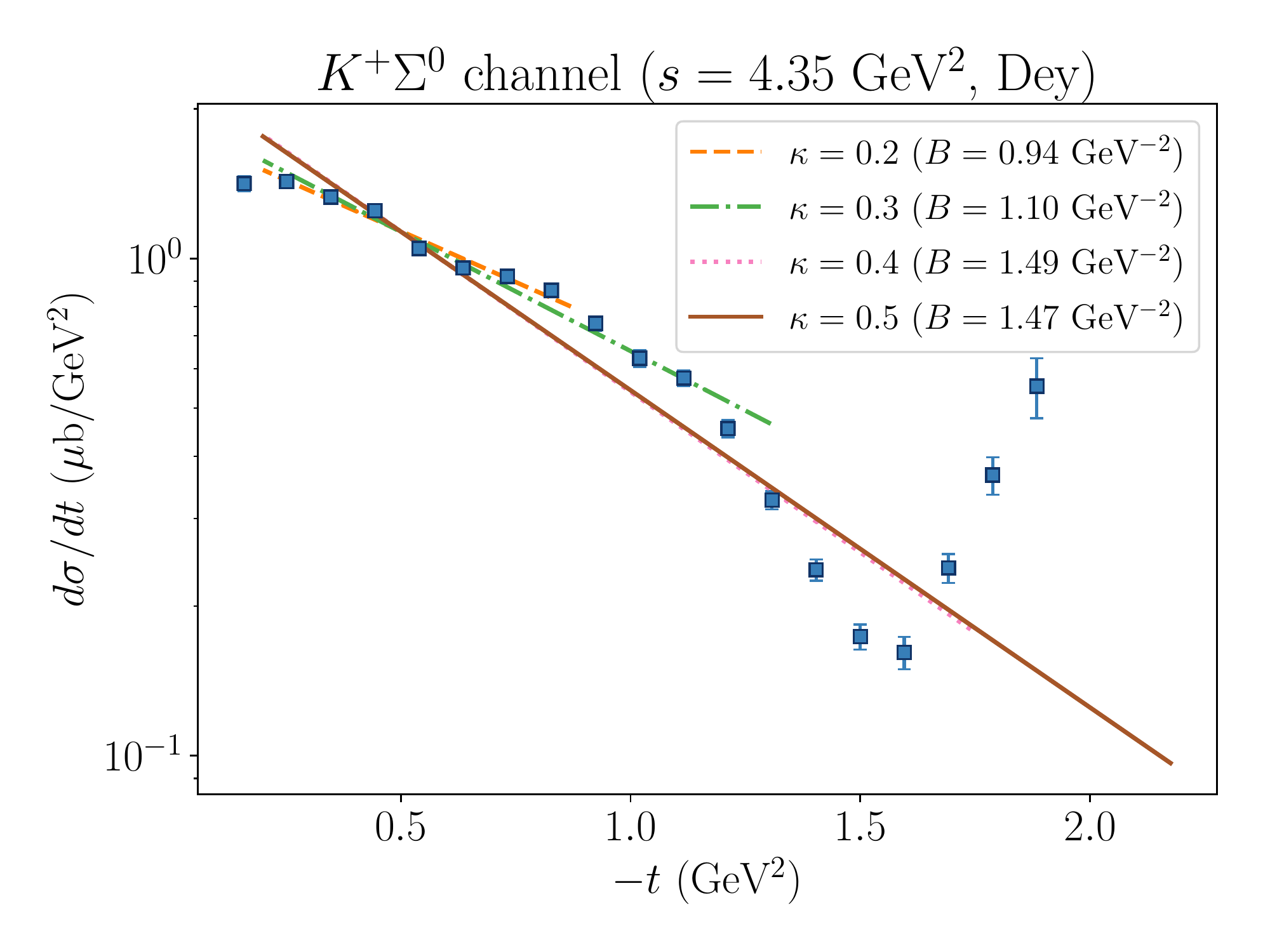}
  \end{subfigure}
  \caption{
    The differential cross section $\frac{d\sigma}{dt}$ versus $-t$ for the
    $K^+\Lambda$ and $K^+\Sigma^0$ channels at
    $s = 4.35$~GeV$^2$,
    for several values of $\kappa=-t_{\mathrm{max}}/s$.
    Fits were made to the plotted CLAS data from
    (top) Bradford {\sl et al.}~\cite{Bradford:2005pt},
    (middle) McCracken {\sl et al.}~\cite{McCracken:2009ra},
    and (bottom) Dey {\sl et al.}~\cite{Dey:2010hh}.
  }
  \label{fig:MKFit}
\end{figure}

However, at smaller $s$, there is not such a visibly clear delineation
between the region where Eq.~(\ref{eqn:AeBt}) works and where it fails.
For instance, in Fig.~\ref{fig:MKFit},
one sees that the differential cross section for
photoproduction of the $K^+\Lambda$ and $K^+\Sigma^0$ channels
at $s=4.35$~GeV$^2$ is not quite a straight line on a log-scaled plot,
but is slightly convex-down.
This means that the extracted value of $B$ will
vary with the region of $-t$ that is fit.
This dependence of $B$ on the fitted $t$ region has been observed before,
especially for photoproduction of $\phi(1020)$
(see {\sl e.g.},~\cite{Bauer:1977iq,Behrend:1978ik,Seraydaryan:2013ija,Dey:2014tfa}).
Accordingly, it is necessary to develop a quantitative scheme
for determining the fit region that is to be used,
and for determining the systematic uncertainty
in $B$ originating from this decision.
Moreover, this scheme should help us determine the minimum value of $s$
to which Eq.~(\ref{eqn:AeBt}) can be fruitfully applied.

Before proceeding, it is necessary to additionally exclude
data with sufficiently small $-t$ from the fit since,
as can be seen in both Figs.~\ref{fig:highs} and \ref{fig:MKFit},
the logarithmic slope of the differential cross section
flattens out and even falls when $-t$ is close to zero.
To this end, we have fit the function $Ate^{Bt}$ to low-$t$ data,
looking for the critical point $t_0 = -B^{-1}$
corresponding to the maximum cross section.
Since the fit function of Eq.~(\ref{eqn:AeBt}) is
strictly monotonically decreasing,
we use the critical point for each $s$ bin as
the $t_{\mathrm{min}}$ for this bin.
On average, $-t_{\mathrm{min}}$ is about $0.25$~GeV$^2$.

To exclude large $-t$ data from the fit, we define a quantity
\begin{equation}
  \kappa = \frac{-t_{\mathrm{max}}}{s}
  \label{eqn:kappa}
\end{equation}
and study the dependence of the extracted $B$ on the value of $\kappa$.
For example, in Fig.~\ref{fig:MKFit},
the differential cross section of the $K^+\Lambda$ and $K^+\Sigma^0$ channels
were fit to the Eq.~(\ref{eqn:AeBt}) within the region
$0.25~\mathrm{GeV}^2 < -t < \kappa s$,
using multiple values of $\kappa$.
The fits in this figure were performed by minimizing $\chi^2$.
The extracted $B$ values for the $K^+\Lambda$ channel
seem to be robust against variation of $\kappa$,
but this is due predominantly to
the relatively larger uncertainties
in the high $-t$ data than in the low $-t$ data,
a difference which causes the low $-t$ data to
overwhelmingly dominate the least-$\chi^2$ fit.
Nonetheless, it is clear that the higher $-t$ data
($-t \ge 2$~GeV$^2$) in
Fig.~\ref{fig:MKFit}
do not follow the linear trend that the fit suggests,
and high $\kappa$ in this sense produces a bad fit.

\begin{figure}
  \centering
  \begin{subfigure}[b]{.48\textwidth}
    \centering
    \includegraphics[width=\textwidth]{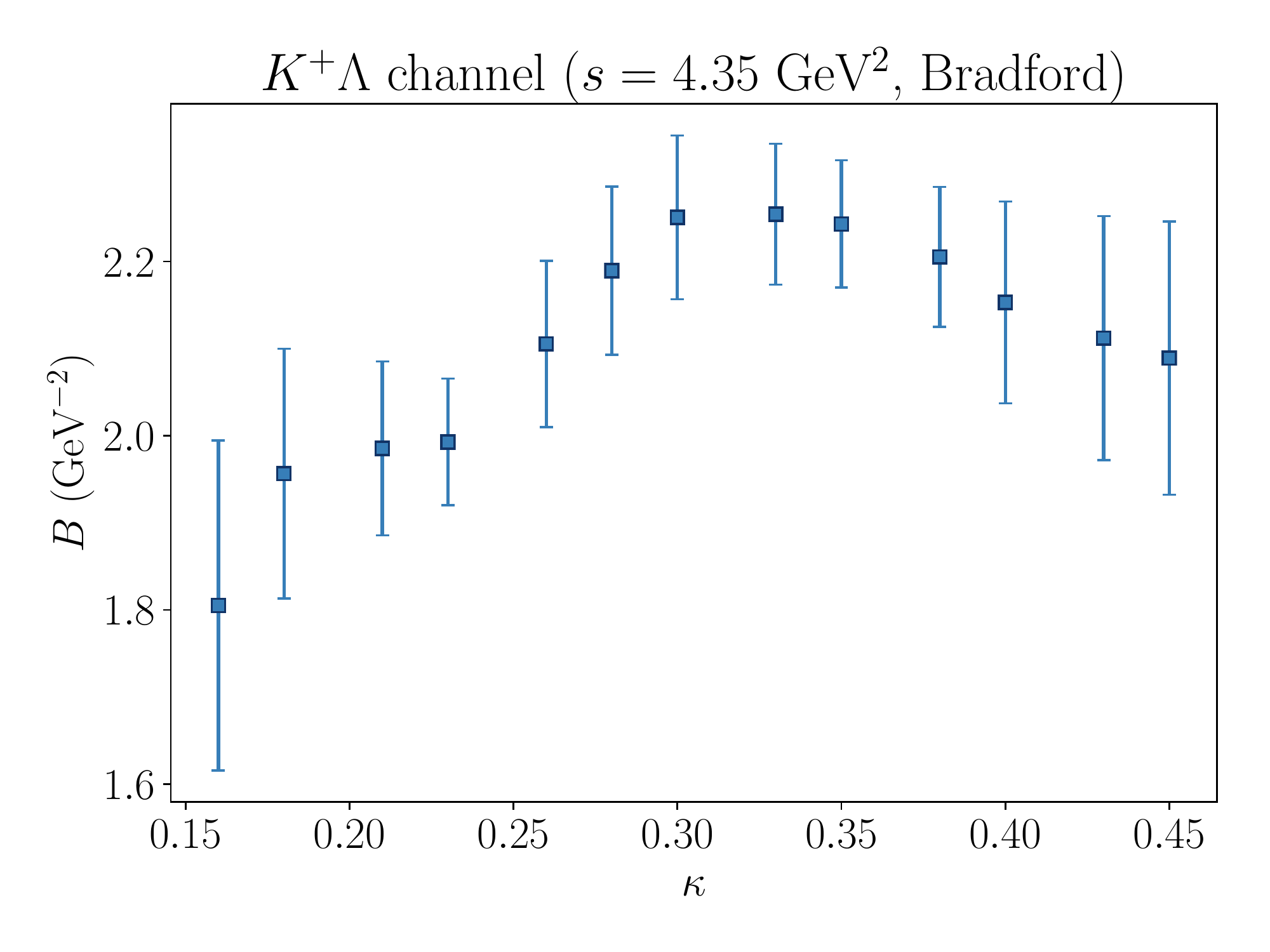}
  \end{subfigure}
  \begin{subfigure}[b]{.48\textwidth}
    \centering
    \includegraphics[width=\textwidth]{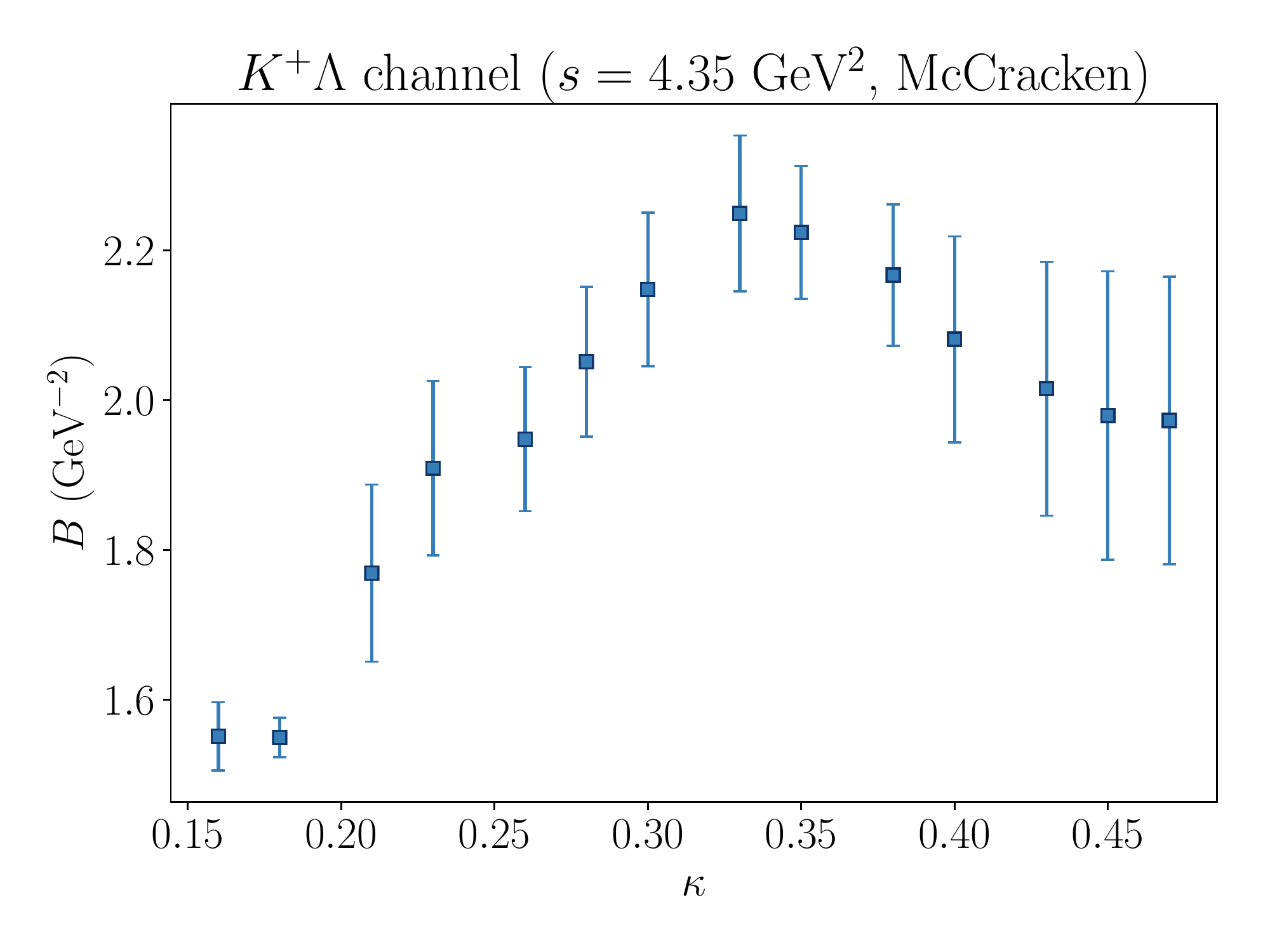}
  \end{subfigure}
  \begin{subfigure}[b]{.48\textwidth}
    \centering
    \includegraphics[width=\textwidth]{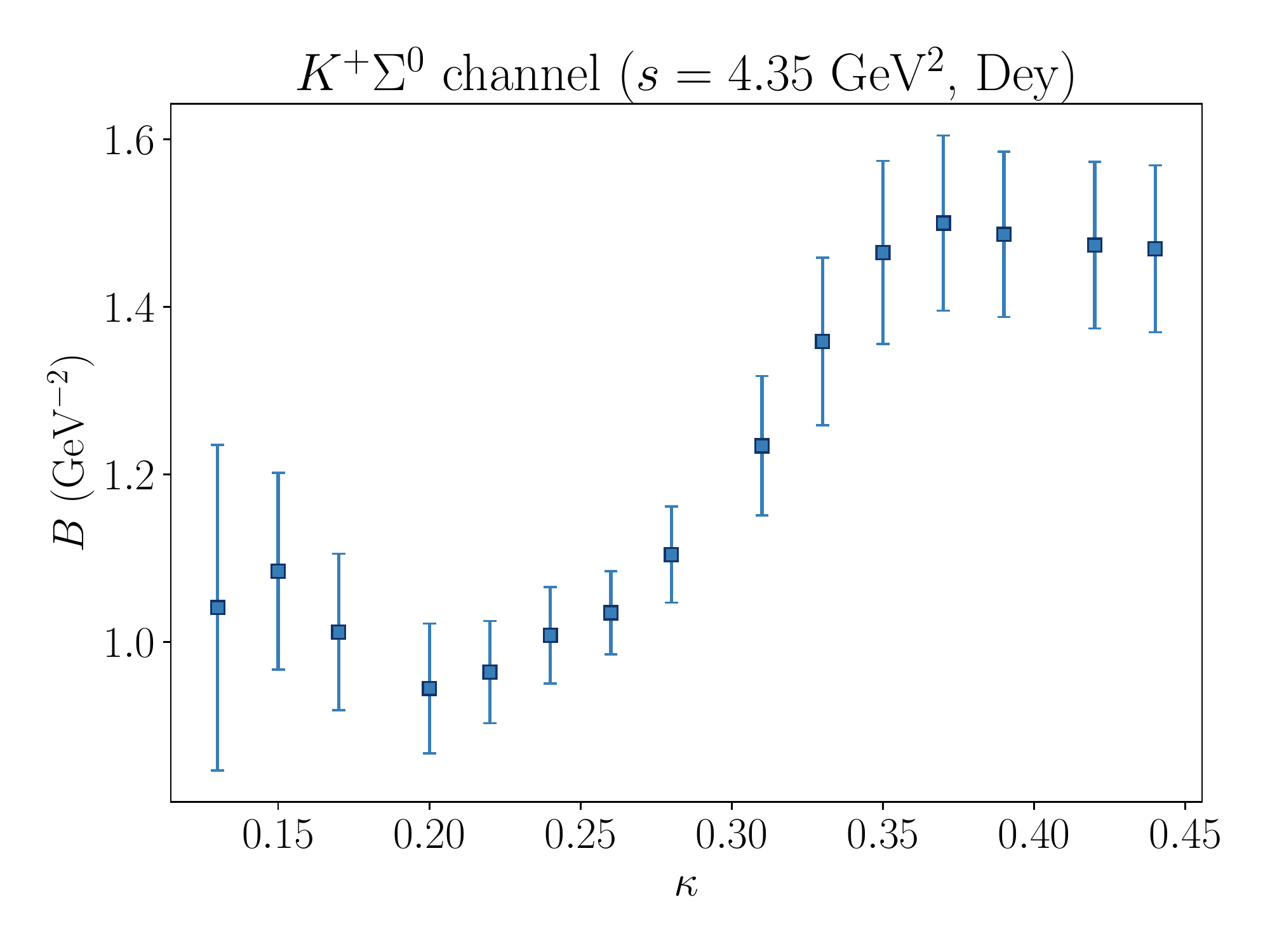}
  \end{subfigure}
  \caption{
    The extracted slope factor $B$ versus $\kappa$ for the
    $K^+\Lambda$ and $K^+\Sigma^0$ channels
    at $s = 4.35$~GeV$^2$.
    The uncertainties in $B$ are due to quality of the fits.
    CLAS data from
    (top) Bradford {\sl et al.}~\cite{Bradford:2005pt},
    (middle) McCracken {\sl et al.}~\cite{McCracken:2009ra},
    and (bottom) Dey {\sl et al.}~\cite{Dey:2010hh}
    were used in these extractions.
  }
  \label{fig:BvK}
\end{figure}

One trend we can see in Fig.~\ref{fig:MKFit} is that
the differential cross section begins falling more steeply as
$-t$ increases from $0$ through around $1$~GeV$^2$ or so,
and at sufficiently high $-t$ the fall-off stops
and the differential cross section begins rising again.
Accordingly, we expect that the extracted $B$,
taken as a function of $\kappa$,
should rise with $\kappa$ to some maximum value before falling.
In light of this expectation,
we have plotted the dependence of $B$ on $\kappa$
for both the $K^+\Lambda$ and $K^+\Sigma^0$ channels in Fig.~\ref{fig:BvK}.
The expectation is indeed met.

\begin{figure}
  \centering
  \begin{subfigure}[b]{.48\textwidth}
    \centering
    \includegraphics[width=\textwidth]{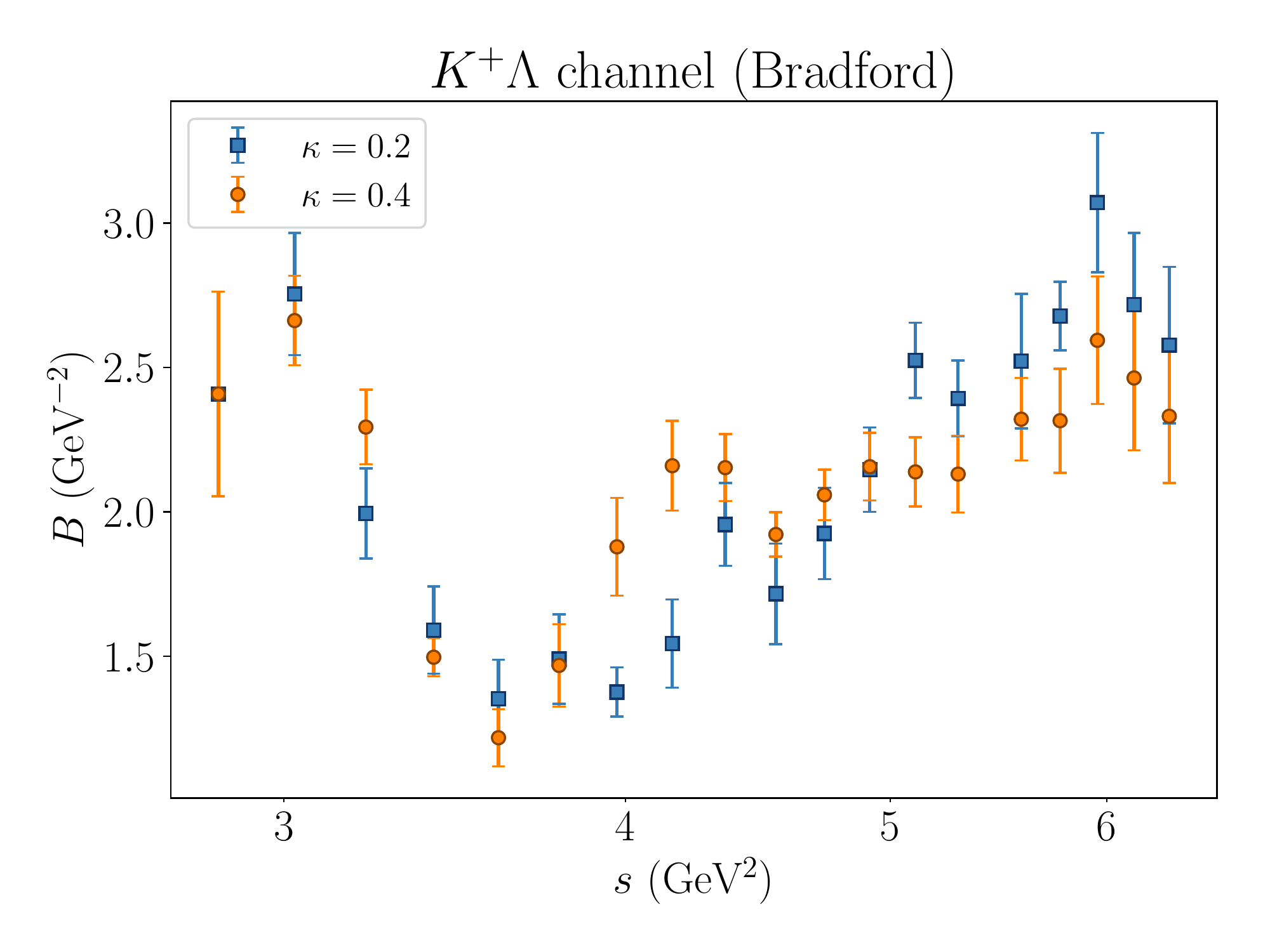}
  \end{subfigure}
  \begin{subfigure}[b]{.48\textwidth}
    \centering
    \includegraphics[width=\textwidth]{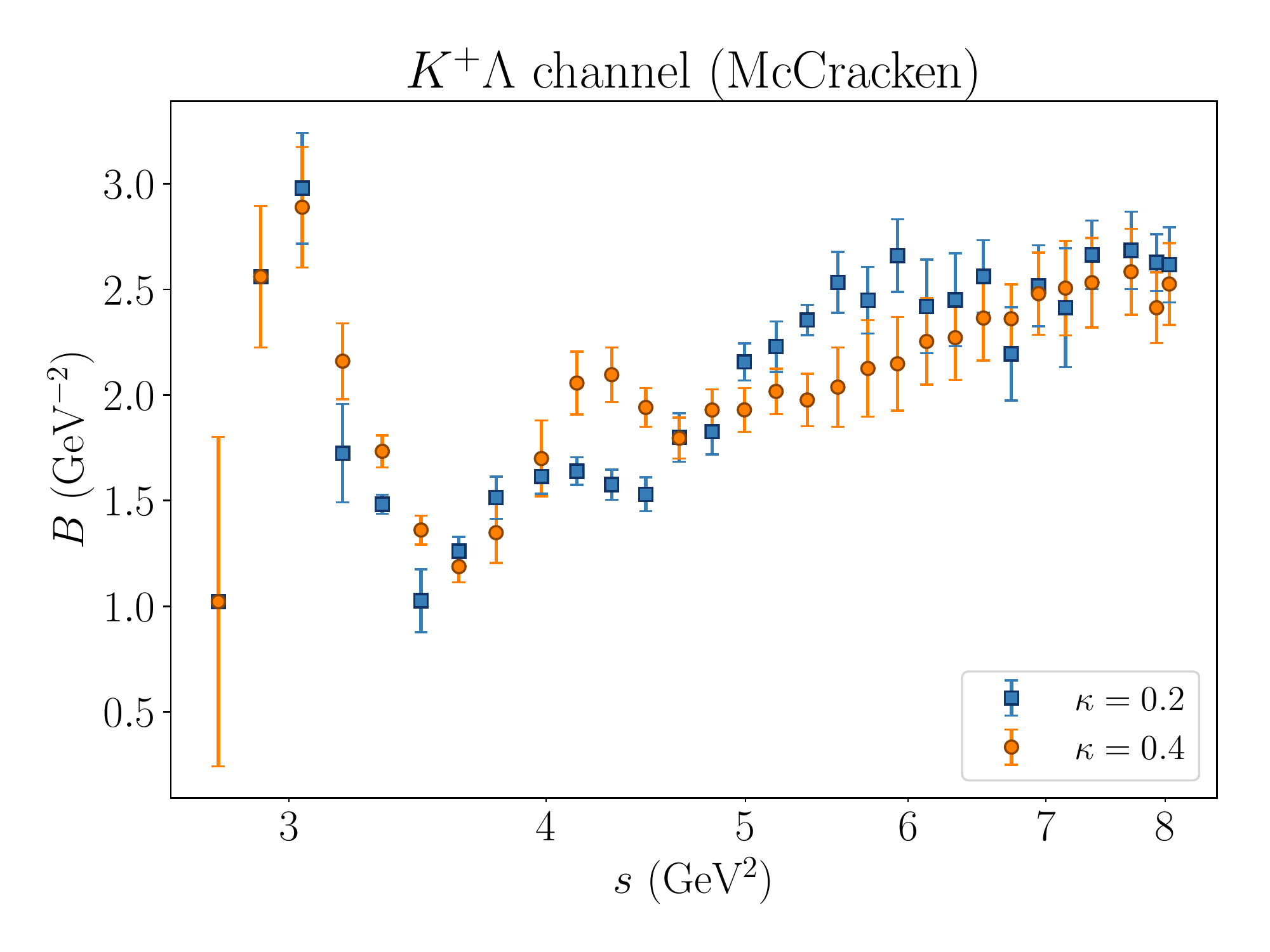}
  \end{subfigure}
  \begin{subfigure}[b]{.48\textwidth}
    \centering
    \includegraphics[width=\textwidth]{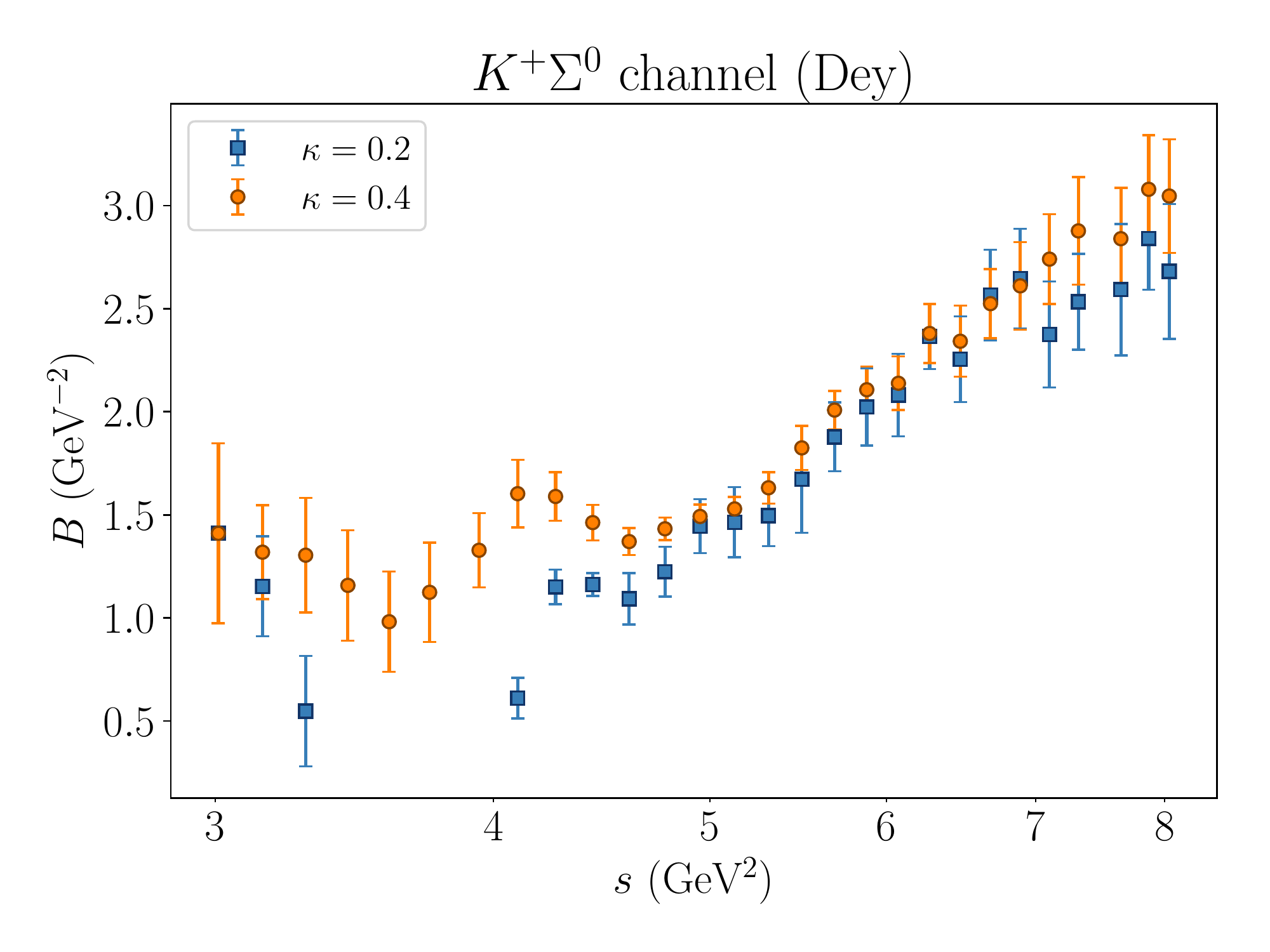}
  \end{subfigure}
  \caption{
    The $t$-slope $B$ versus $s$ for the
    $K^+\Lambda$ and $K^+\Sigma^0$ channels,
    extracted from a fit to Eq.~(\ref{eqn:AeBt}).
    Two values of $\kappa$ (as defined in Eq.~(\ref{eqn:kappa}))
    were used for the fits.
    CLAS data from
    (top) Bradford {\sl et al.}~\cite{Bradford:2005pt},
    (middle) McCracken {\sl et al.}~\cite{McCracken:2009ra},
    and (bottom) Dey {\sl et al.}~\cite{Dey:2010hh}
    were used.
    To improve readability of the plots, every fourth data
    point was plotted.
  }
  \label{fig:BvS}
\end{figure}

Next, we look at the $s$ dependence of $B$.
Since the choice of $\kappa$ will affect the extracted $B$ value,
we study the $s$ dependence at two different values of $\kappa$,
specifically $\kappa=0.2$ and $0.4$.
We plot several of the extracted $B$ values against $s$ in Fig.~\ref{fig:BvS}.
Since the CLAS data is very finely binned,
we have plotted every fourth data point 
so that the plots are readable.
For the extractions from the more recent CLAS publications from
McCracken {\sl et al.}~\cite{McCracken:2009ra}
and Dey {\sl et al.}~\cite{Dey:2010hh},
in Fig.~\ref{fig:BvS},
the extracted $B$ values for the two different $\kappa$ values fall
within each other's error bars at $s > 5$~GeV$^2$.

It has previously been observed
({\sl e.g.}\ in~\cite{Guidal:1997hy})
that Regge theory's expected domain of validity is $s \gtrsim 5$~GeV$^2$,
and the relative robustness of the $B$ extractions from the McCracken
and Dey values in this domain is compatible with this previous knowledge.
However, there is less consistency between the $\kappa$ values
for the Bradford~{\sl et al.}~data~\cite{Bradford:2005pt},
thus prompting the question of which $\kappa$ value is most appropriate.

We next investigate how to determine the best value of $\kappa$ to use,
and to estimate the systematic uncertainty in $B$
resulting from that choice.
To proceed,
we make use of the fact that (as in Fig.~\ref{fig:BvK})
$B$ as a function of $\kappa$ reaches a maximum value,
corresponding to a range of $-t$ in which the cross section
most behaves like a steeply-falling exponential.
The $\kappa$ which maximizes $B$ is chosen as a ``central'' value,
and is denoted $\kappa_0$.
Two lower and higher values of $\kappa$
are chosen which produce different $t$-slopes
(denoted $\kappa_{-2}$, $\kappa_{-1}$, $\kappa_1$, and $\kappa_2$),
and the standard deviation of the $t$-slopes produced by these $\kappa$ values is
taken as the systematic ``$\kappa$ choice'' uncertainty.
Namely, we define
\begin{equation}
  \sigma^{(\mathrm{sys})}_B = \sqrt{
    \sum_{i=-2}^2 \left(B(\kappa_i)\right)^2
    - \left(\sum_{i=-2}^2 B(\kappa_i)
    \right)^2
  }
  \label{eqn:SysErr}
  .
\end{equation}

\begin{figure}
  \centering
  \begin{subfigure}[b]{.48\textwidth}
    \centering
    \includegraphics[width=\textwidth]{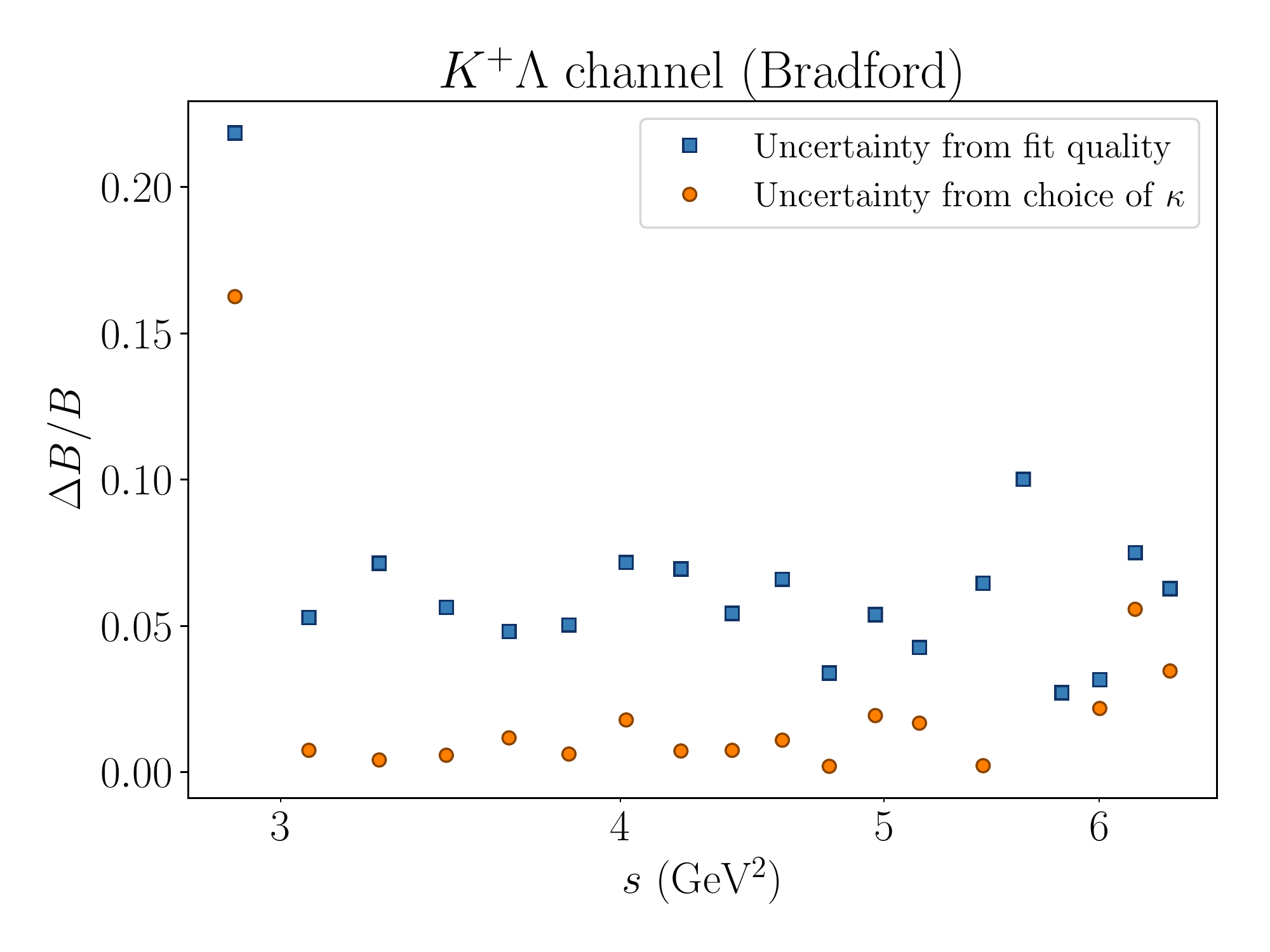}
  \end{subfigure}
  \begin{subfigure}[b]{.48\textwidth}
    \centering
    \includegraphics[width=\textwidth]{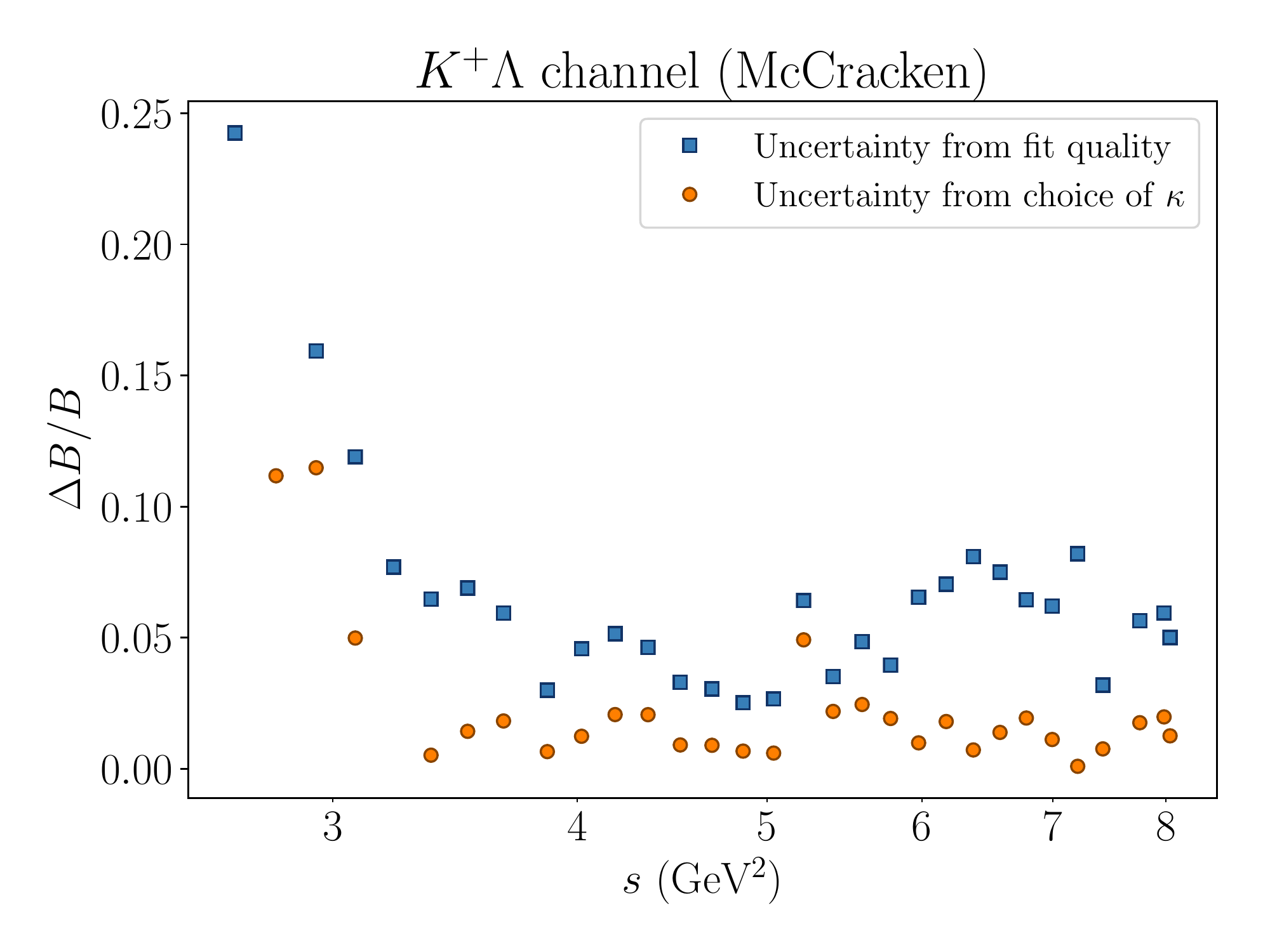}
  \end{subfigure}
  \begin{subfigure}[b]{.48\textwidth}
    \centering
    \includegraphics[width=\textwidth]{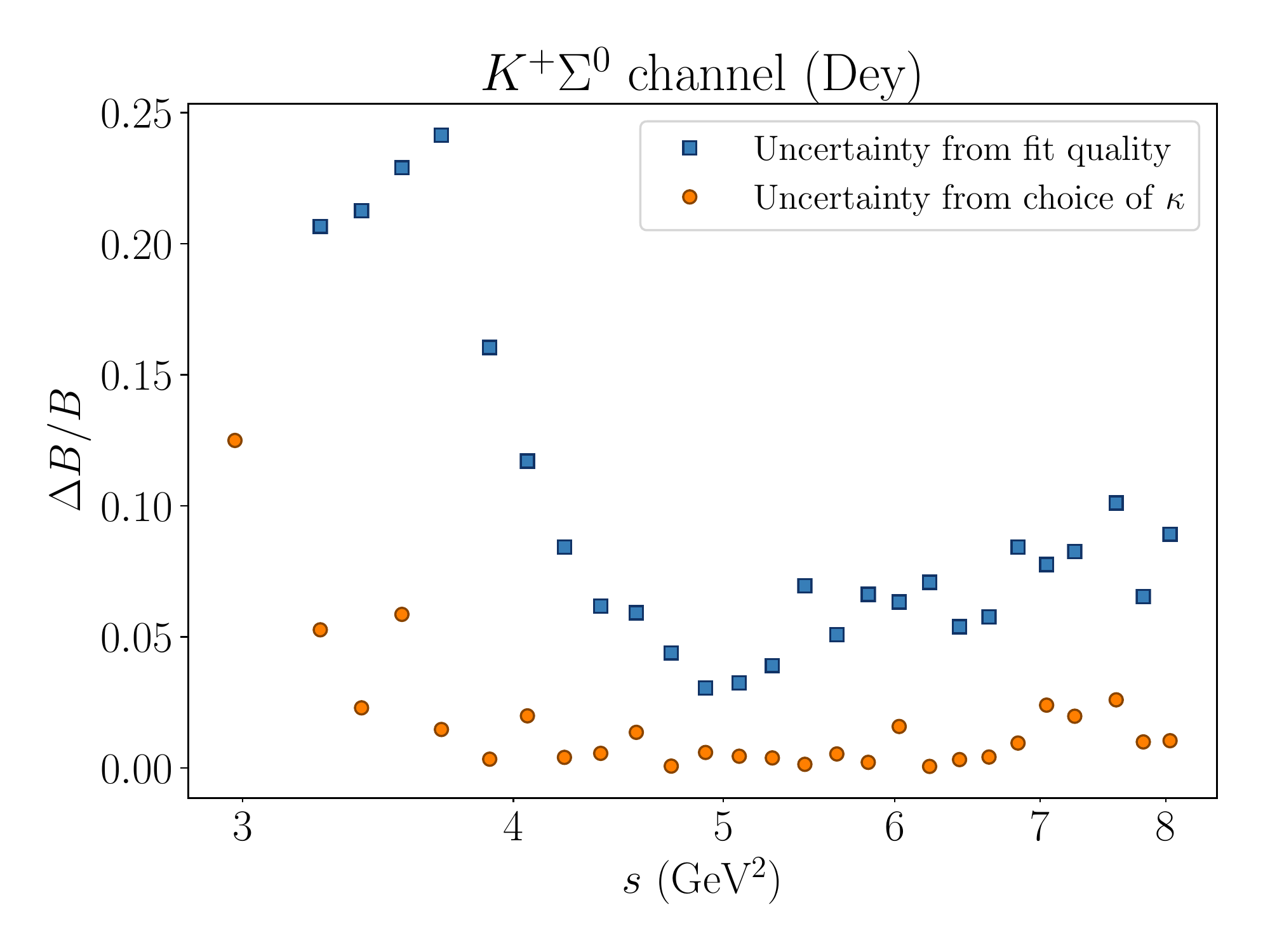}
  \end{subfigure}
  \caption{
    The relative uncertainty in $B$ versus $s$, for two sources of uncertainty:
    the uncertainty due to the quality of the fit,
    and the ``systematic'' uncertainty due to the choice of $\kappa$.
    CLAS data from
    (top) Bradford {\sl et al.}~\cite{Bradford:2005pt},
    (middle) McCracken {\sl et al.}~\cite{McCracken:2009ra},
    and (bottom) Dey {\sl et al.}~\cite{Dey:2010hh}
    were used in extracting the $B$ values.
    To improve readability of the plots, every fourth data
    point was plotted.
  }
  \label{fig:SysErr}
\end{figure}

In Fig.~\ref{fig:SysErr},
we have plotted the relative uncertainty in the $t$-slope
due to both the quality of the fits,
and the systematic uncertainty due to the choice of $\kappa$.
It is worth noting that the $\kappa$ choice uncertainty
(for the central value we have chosen)
is significantly smaller than the uncertainty due to quality of the fits.

\begin{figure*}
  \centering
  \begin{subfigure}[b]{\textwidth}
    \centering
    \includegraphics[width=\textwidth]{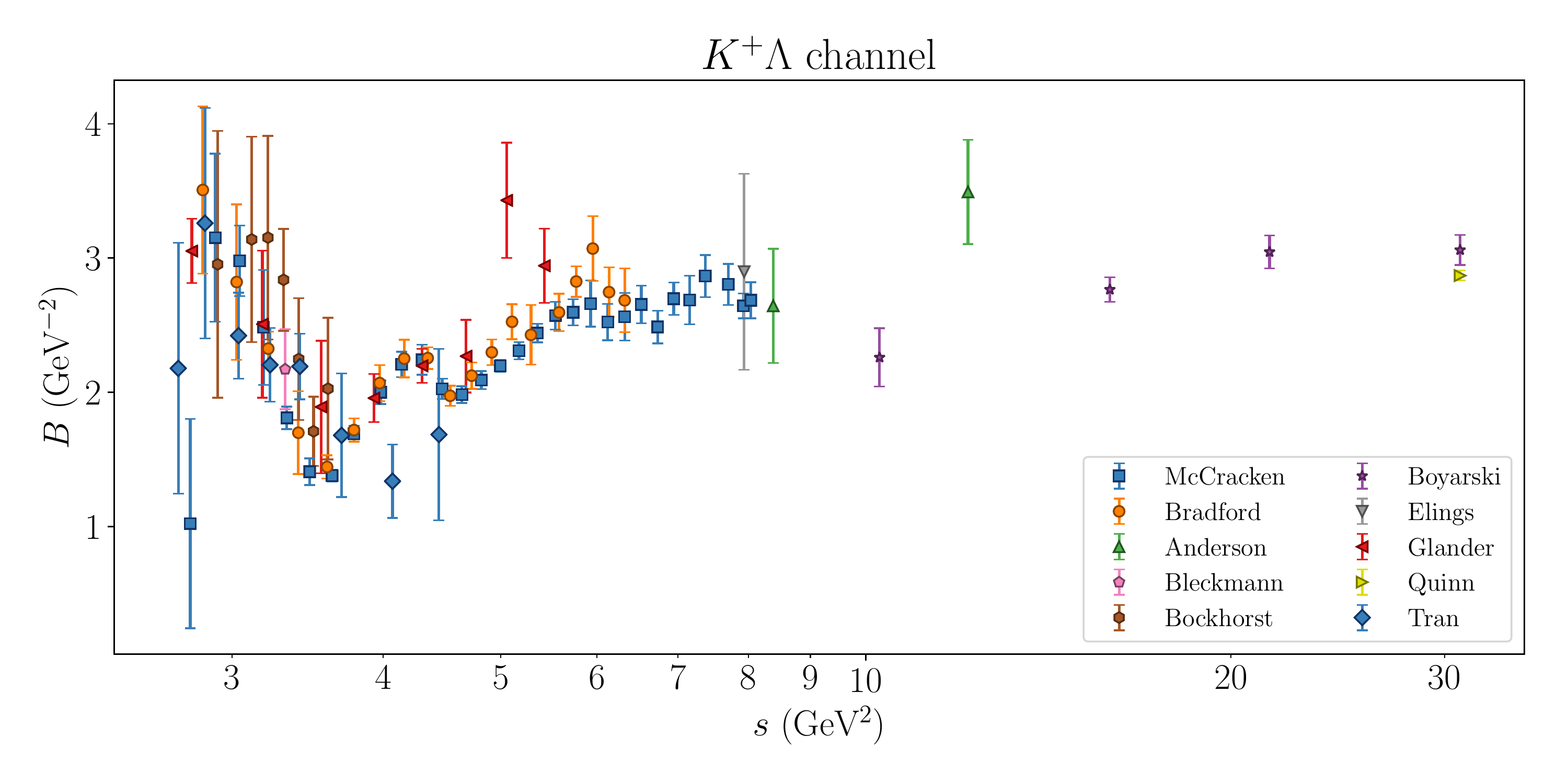}
  \end{subfigure}
  \begin{subfigure}[b]{\textwidth}
    \centering
    \includegraphics[width=\textwidth]{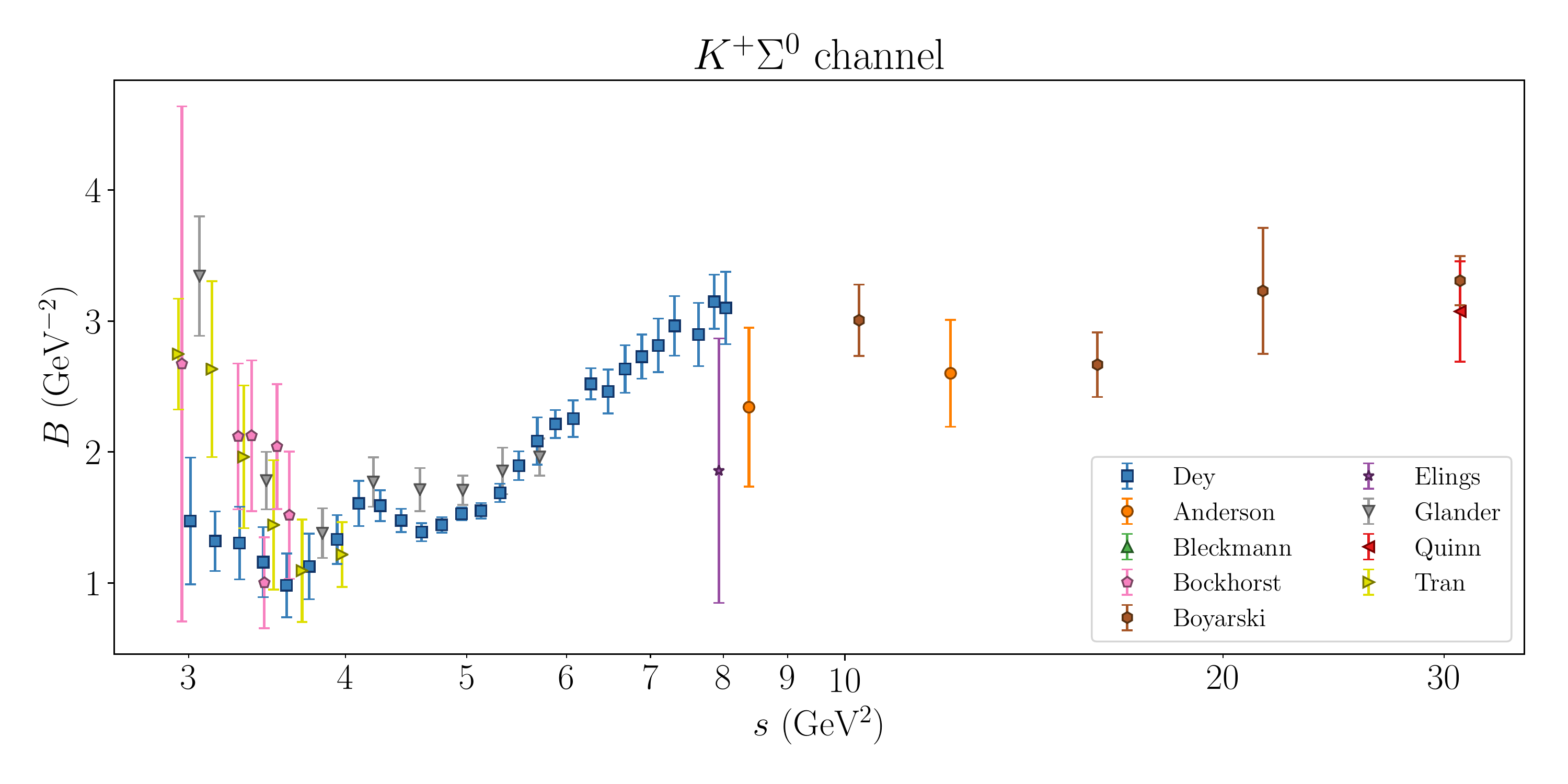}
  \end{subfigure}
  \caption{The $t$-slope $B$ versus $s$ for the
    (top) $K^+\Lambda$ and
    (bottom) $K^+\Sigma^0$ channels.
    The $t$-slopes were extracted by finding
    the $\kappa$ value which produces the maximum $B$ value.
    The data included in these plots is
    a subset of the world data in the kinematic range of
    interest~\cite{Elings:1967af,Boyarski:1970yc,
      Bleckmann:1970kb,Anderson:1976ph,Quinn:1979zp,Bockhorst:1994jf,
      Tran:1998qw,Glander:2003jw,Bradford:2005pt,McCracken:2009ra,Dey:2010hh}.
    For the finely-binned CLAS data, every fourth data point was plotted
    in order to make the plot readable.
  }
  \label{fig:Collabs}
\end{figure*}

In Fig.~\ref{fig:Collabs}, we use the ``maximum $B$'' scheme
for determining the slope factor,
and have plotted the extracted slope factors against $s$.
We have included the world data for small-angle photoproduction of both
$K^+\Lambda$ and $K^+\Sigma^0$ from a proton target in this figure.
For the finely-binned
CLAS data~\cite{Bradford:2005pt,McCracken:2009ra,Dey:2010hh}, 
we have plotted every fourth data point in order to keep the plot readable.
In these plots, the newer CLAS data from McCracken~\cite{McCracken:2009ra}
and Dey~\cite{Dey:2010hh} seem to fall roughly along straight lines
in the region $5 < s < 8.1$~GeV$^2$
when the $x$-axis is log-scaled.
This is in accordance with the prescription of Eq.~(\ref{eqn:B}).

\begin{figure*}
  \centering
  \includegraphics[width=\textwidth]{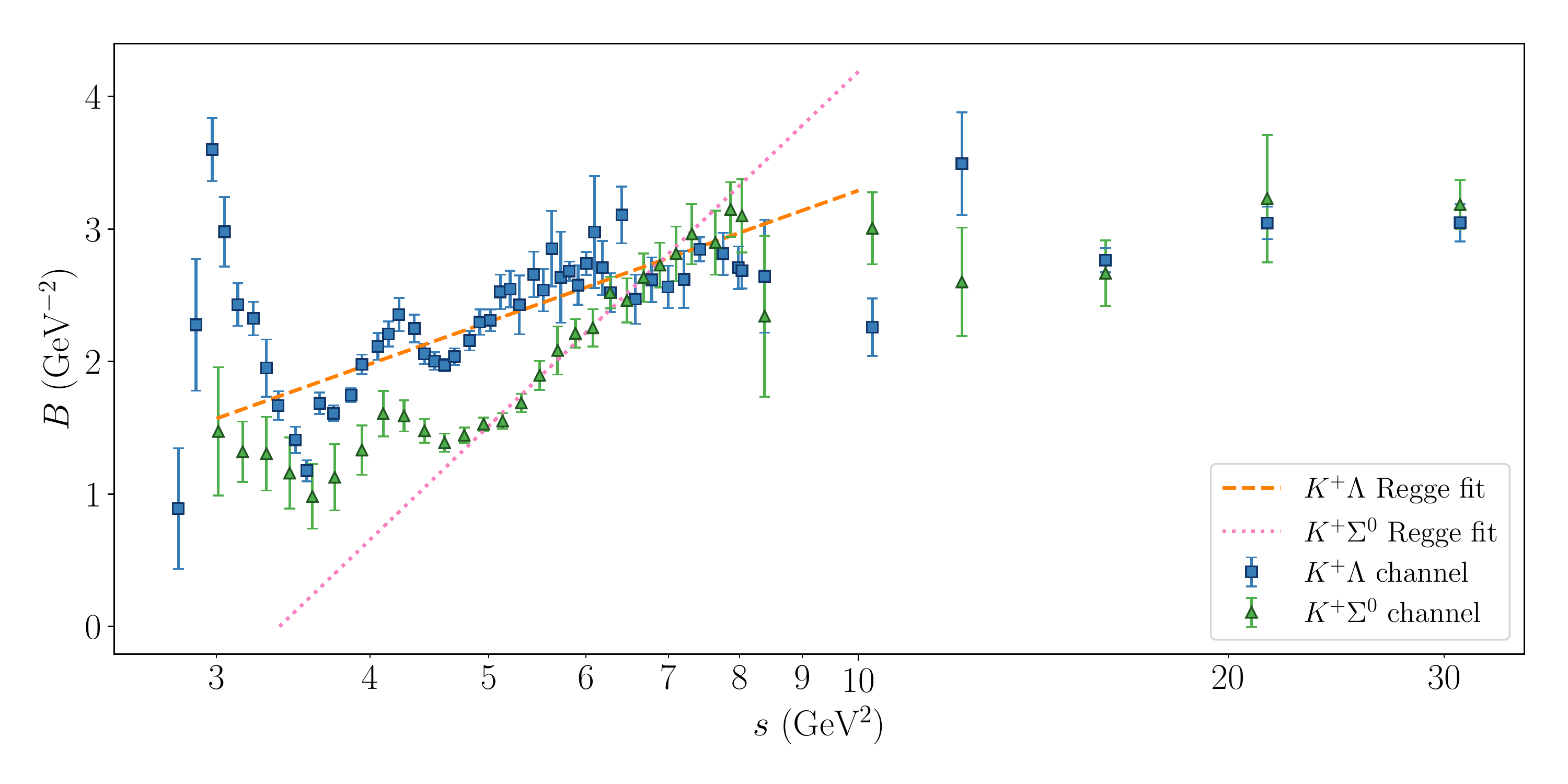}
  \caption{
    The $t$-slope $B$ for both the $K^+\Lambda$ and $K^+\Sigma^0$ channels,
    determined using the maximum $B$ scheme.
    The $s < 8.1$~GeV$^2$ points are CLAS data from
    McCracken~\cite{McCracken:2009ra} and Dey~\cite{Dey:2010hh},
    for which every fourth point was plotted to improve readability.
    SLAC data~\cite{Boyarski:1970yc,Anderson:1976ph,Quinn:1979zp}
    at $s > 8.1$~GeV$^2$ are included for comparison.
    The lines are Regge fits to the recent CLAS data
    in the region $5 < s < 8.1$~GeV$^2$ (see text).
  }
  \label{fig:MaxB}
\end{figure*}

Since Fig.~\ref{fig:Collabs} seems to vindicate Eq.~(\ref{eqn:B}),
we proceed to extract the parameters $\alpha^\prime$ and $s_0$
by fitting Eq.~(\ref{eqn:B}) to the recent CLAS data from
McCracken {\sl et al.}~\cite{McCracken:2009ra}
and Dey {\sl et al.}~\cite{Dey:2010hh}.
We perform this fit in the region
$5$~GeV$^2 < s < 8.1$~GeV$^2$ for both the $K^+\Lambda$
and $K^+\Sigma^0$ channels.
This produces extracted Regge slopes of
\begin{eqnarray}
  \alpha^\prime_{K^+\Lambda} &= 0.60\pm 0.05~\mathrm{GeV}^{-2} \\
  \alpha^\prime_{K^+\Sigma^0} &= 1.93\pm 0.05~\mathrm{GeV}^{-2}
\end{eqnarray}
and extracted $s_0$ values of
\begin{eqnarray}
  s_0^{(K^+\Lambda)} &= 0.72\pm 0.12~\mathrm{GeV}^2 \\
  s_0^{(K^+\Sigma^0)} &= 3.37\pm 0.04~\mathrm{GeV}^2
\end{eqnarray}
Neither of the extracted trajectories in this two-parameter fit
is compatible with one-Reggeon exchange,
although the $K^+\Lambda$ channel comes close.
In fact, it is possible to peform a one-paraeter
fit the $K^+\Lambda$ data with the constraint
$s_0 = 1$~GeV$^2$ with little increase in $\chi^2$ of the fit,
which would give us
\begin{equation}
  \alpha^\prime_{K^+\Lambda} = 0.71~\mathrm{GeV}^{-2}
  .
\end{equation}
This is compatible with pure $K$ Reggeon exchange,
since $\alpha^\prime_{K} = 0.69\pm 0.02$~GeV$^{-2}$.
On the other hand, the trajectory for the $K^+\Sigma^0$ channel far exceeds
the trajectories of both the $K$ and the $K^*$,
and it is not possible to contrain $s_0 = 1$~GeV$^2$ for the $K^+\Sigma^0$
channel without increasing $\chi^2$ by an order of magnitude,
suggesting that the $K^+\Sigma^0$ final state
is not produced by exchange of a single Reggeon.

Regarding the extracted $s_0$ values,
we remind the reader that
the parameter $s_0$ is not truly a constant,
but a function of $t$,
which can approximately be written as~\cite{Collins:1977jy}
\begin{equation}
  s_0(t) \approx \frac{\sqrt{
      t^2 - 2t(m_p^2+m_X^2) + (m_p^2-m_X^2)^2
  }}{2}
  \label{eqn:s0}
  ,
\end{equation}
where $m_X=m_\Lambda$ or $m_{\Sigma^0}$, depending on the channel.
Eq.~(\ref{eqn:s0}) gives $s_0(t)\approx 1$~GeV$^2$ when $-t$ is small,
and this is the conventional value of $s_0$ used in
Regge phenomenology~\cite{Collins:1977jy,Guidal:1997hy}.
The value
$s_0^{(K^+\Lambda)} = 0.72\pm 0.12$~GeV$^2$
that we obtained is roughly compatible with this,
but
$s_0^{(K^+\Sigma^0)} = 3.37\pm 0.04$~GeV$^2$
is far in excess of the values $s_0$ will take in the $t$ range of interest.
We interpret this to vindicate both that
the $K^+\Lambda$ channel is produced mostly by $K$ Reggeon exchange,
and that the $K^+\Sigma^0$ channel cannot be explained by
single Reggeon exchange.
Compatibility with Eq.~(\ref{eqn:B}) is indicated, after all,
not just by a logarithmic dependence of $B$ on $s$,
but by the parameters in the equation representing true physical values.

It is further worth noting that the linear trend
followed by the CLAS data in Figs.~\ref{fig:Collabs} and \ref{fig:MaxB}
stops at around $s\approx 8$~GeV$^2$.
At higher $s$, the existing world data is sparse but,
when taken together with the lower-$s$ CLAS data,
is suggestive of a flattening out of $B$.
However, a finer structure that cannot be discerned
with the current world data is also possible.
In either case, the functional form suggested by Eq.~(\ref{eqn:B})
appears to be violated at larger $s$.
It would be prudent to extend the experimental investigation
of $K^+$ meson photoproduction to higher energies
in order to obtain more high-precision data in the $s > 8$~GeV$^2$ region.
This will be possible at Jefferson Lab with the recent 12~GeV upgrade.


\section{Double-exponential Fit}
\label{sec:double}

It was observed in the previous section that
the ``trajectory'' of the $K^+\Sigma^0$ channel
exceeded the expected trajectories for both $K$ and $K^*$ Reggeon exchange,
and we hypothesized that this was due to exchange of multiple Reggeons.
In this section, we offer a preliminary investigation into this possibility
by fitting the cross section to the interfering sum of two trajectories.
In particular, we fit the cross section to
\begin{equation}
  \frac{d\sigma}{dt} =
  \left|A_{K}(s)e^{B_K(s)t/2} + A_{K^*}(s)e^{B_{K^*}(s)/2}\right|^2
  \label{eqn:double}
  .
\end{equation}
Since we are assuming purely diffractive $t$-channel exchanges,
both $A_{K}(s)$ and $A_{K^*}(s)$ should be predominantly imaginary.
Typically, one would introduce a small real part using
$A(s) = (i+\alpha(s))C(s)$,
but we will here neglect the real part,
taking $\alpha(s)=0$.
Since Eq.~(\ref{eqn:double}) can only be used
to fit $A_K(s)$ and $A_{K^*}(s)$ up to a common phase,
we will normalize the phase so that $A_K(s)$ is positive.

We use Eq.~(\ref{eqn:double}) to fit only $A_K(s)$ and $A_{K^*}(s)$.
The slope factors $B_K(s)$ and $B_{K^*}(s)$ used in the fit
are determined using Eq.~(\ref{eqn:B}),
in which we use the ``conventional'' value of $s_0=1$~GeV$^2$ .
We apply the fit to the range
$0.25~\mathrm{GeV}^2 < -t < \kappa(s) s$,
where $\kappa(s)$ are the $\kappa$ values determined previously
through the ``maximum $B$'' scheme.
With these constraints in place,
the magnitudes and relative sign of $A_K(s)$ and $A_{K^*}(s)$
remain to be fit.

\begin{figure}
  \centering
  %
  %
  \begin{subfigure}[b]{.48\textwidth}
    \centering
    \includegraphics[width=\textwidth]{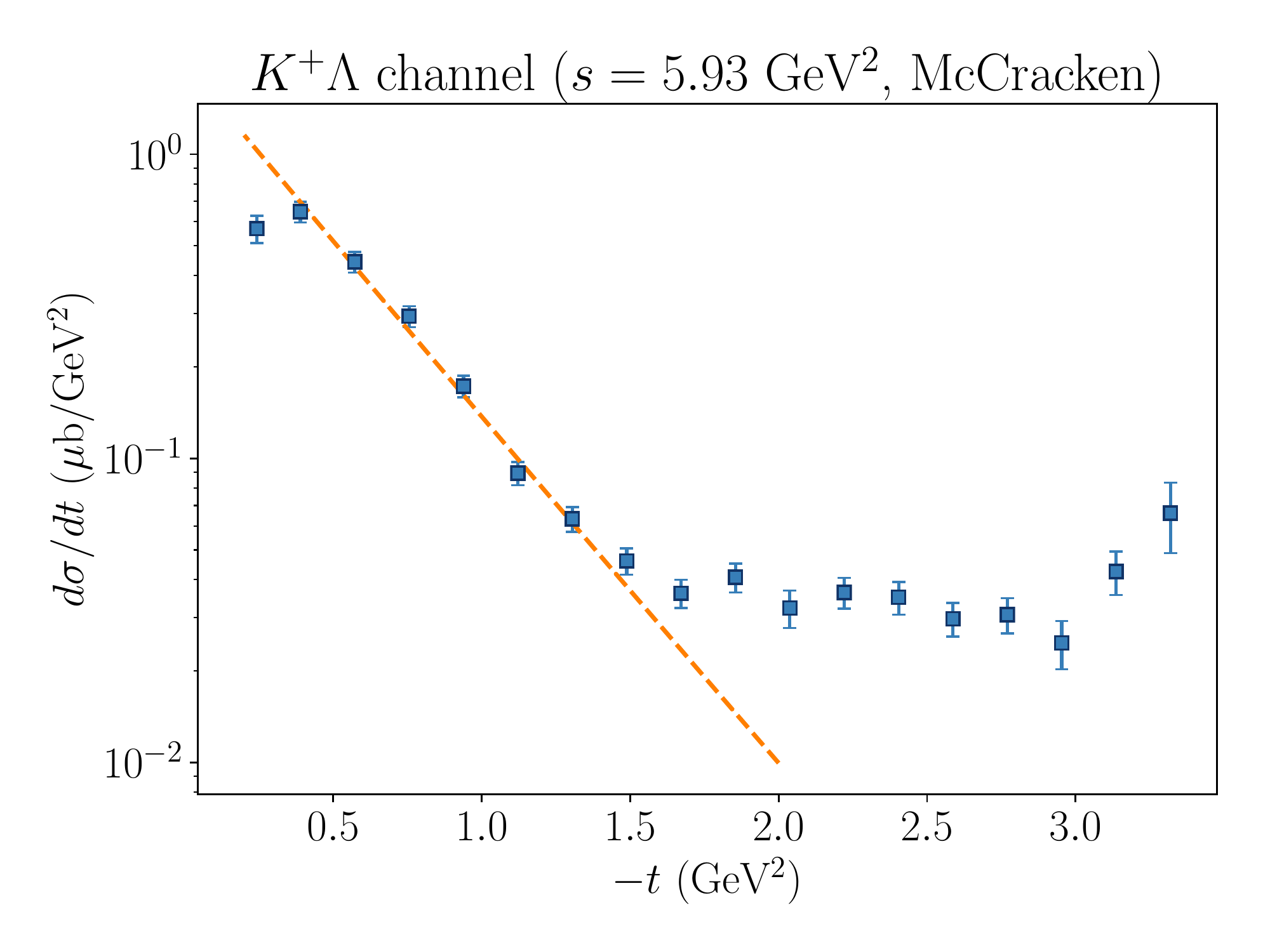}
  \end{subfigure}
  \begin{subfigure}[b]{.48\textwidth}
    \centering
    \includegraphics[width=\textwidth]{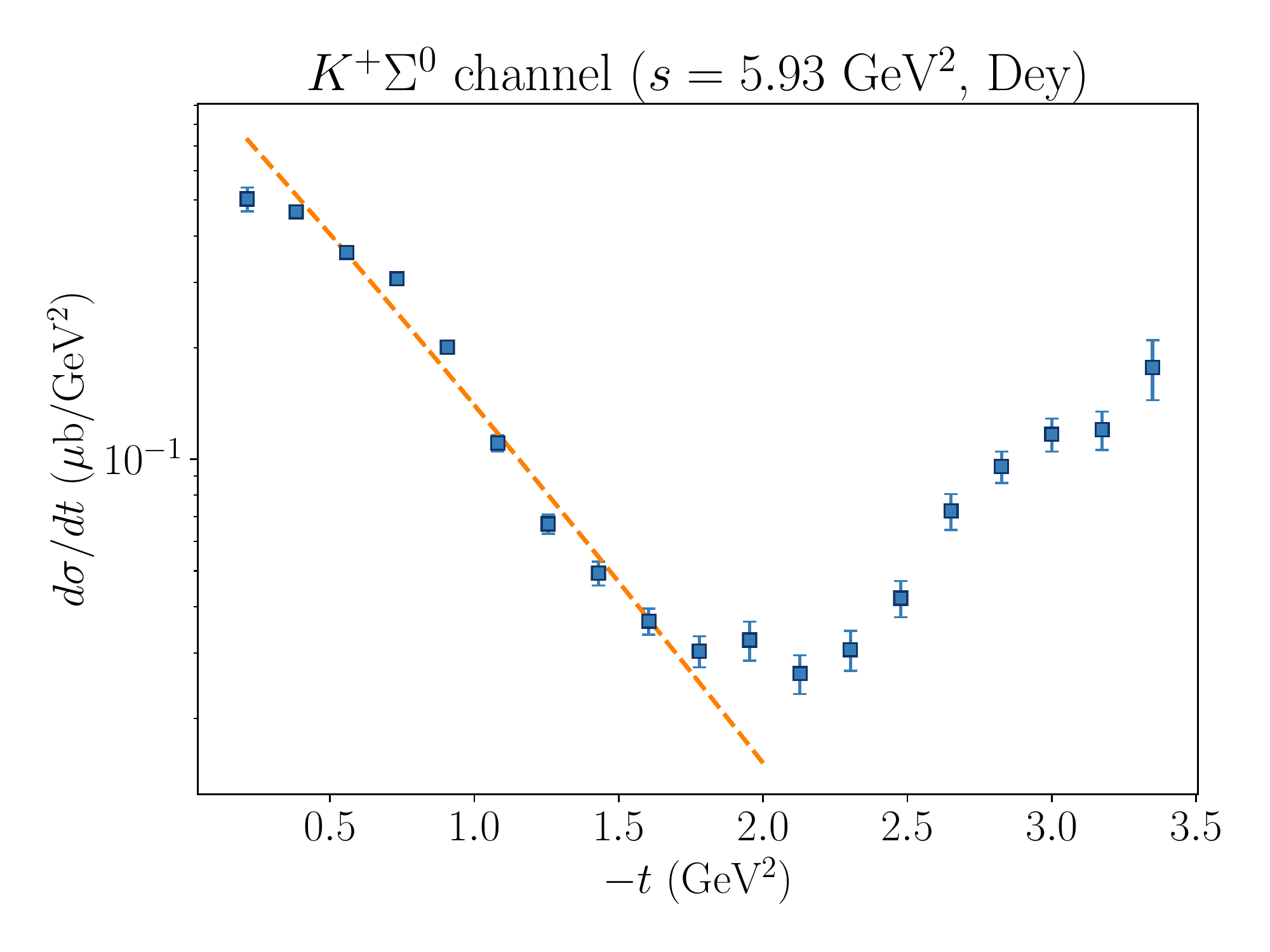}
  \end{subfigure}
  \caption{
    The differential cross section $\frac{d\sigma}{dt}$ versus $-t$
    for the $K^+\Lambda$ and $K^+\Sigma^0$ channels
    at $s = 5.93$~GeV$^2$,
    along with the double-exponential fit to Eq.~(\ref{eqn:double}).
    CLAS data from
    (top) McCracken {\sl et al.}~\cite{McCracken:2009ra},
    and (bottom) Dey {\sl et al.}~\cite{Dey:2010hh}
    were used in these fits.
  }
  \label{fig:DblFit}
\end{figure}

Examples of double-exponential fits are given in Fig.~\ref{fig:DblFit}.
The fits are of variable quality,
with improved description of the data at higher $s$.
We expect the fits to be imperfect because of
several approximations that have been made, including
(1) that all of the $t$-dependence of the differential cross section
comes from the Regge trajectory $\alpha(t)$, and
(2) that the reaction $\gamma p \rightarrow K^+ \Lambda (\Sigma^0)$
proceeds entirely through $t$-channel exchange.
A more complete description including other sources of $t$ dependence
(such as the $t$-dependence of $s_0$)
as well as resonance contributions to the cross section,
such as found in~\cite{Guidal:1997hy,Corthals:2005ce,Corthals:2006nz},
would likely describe the data better.
However, the fits in Fig.~\ref{fig:DblFit} are improvements upon
the simple exponential fits that can be seen in Fig.~\ref{fig:MKFit}.

The purpose of this double-exponential fit is to test the hypotheses that,
in the region $5 < s < 8.1$~GeV$^2$,
(1) the $K^+\Lambda$ channel occurs almost entirely
due to $K$ Reggeon exchange, and
(2) the $K^+\Sigma^0$ channel occurs
due to a mixture of $K$ and $K^*$ Reggeon exchange.
Accordingly, we have applied this fit to
the recent CLAS data in the region $5 < s < 8.1$~GeV$^2$,
and to world data at higher $s$ to check whether
the Reggeon exchange contributions do change around $s\approx 8$~GeV$^2$.

\begin{figure*}
  \centering
  \begin{subfigure}[b]{\textwidth}
    \centering
    \includegraphics[width=\textwidth]{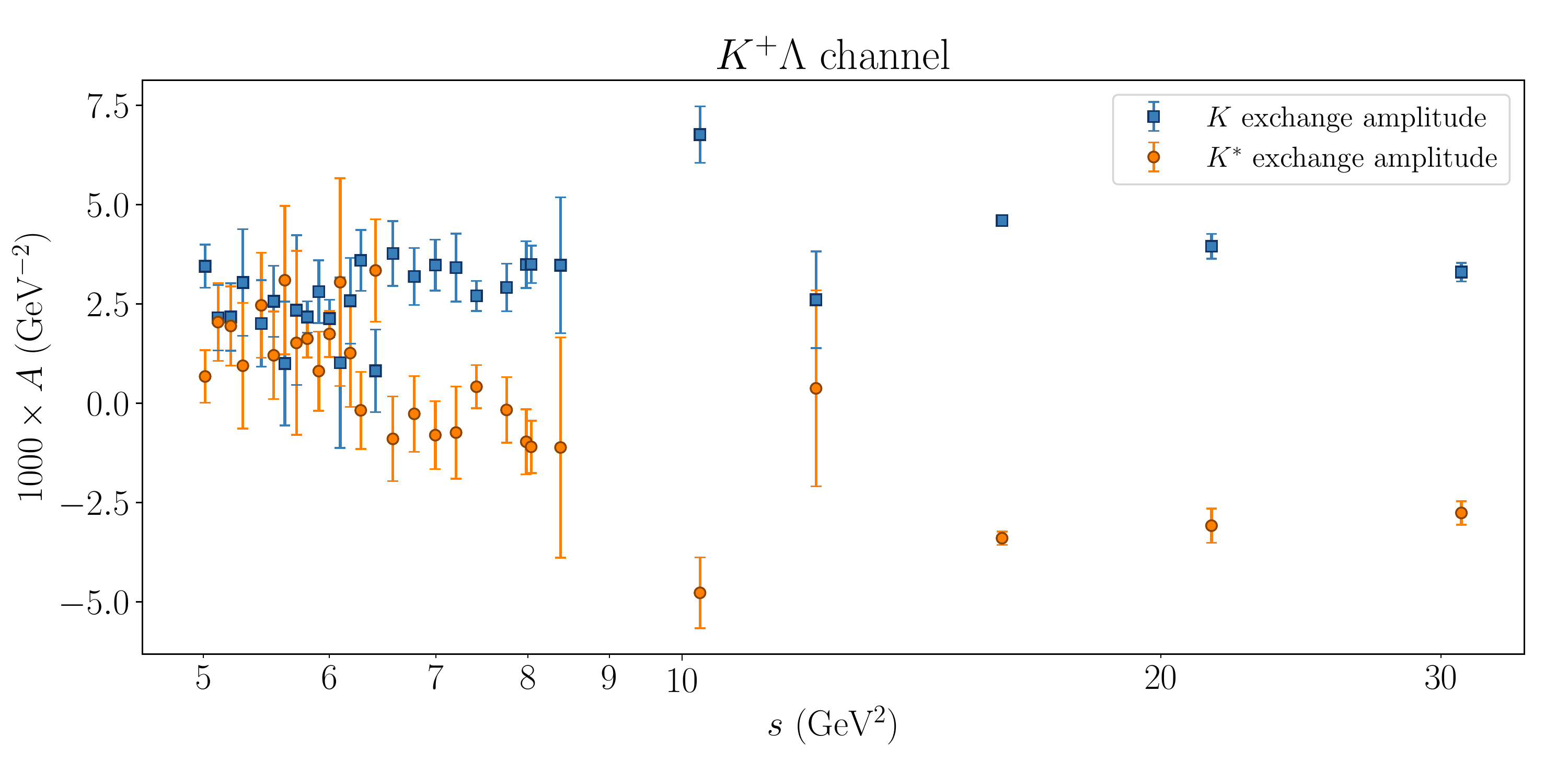}
  \end{subfigure}
  \begin{subfigure}[b]{\textwidth}
    \centering
    \includegraphics[width=\textwidth]{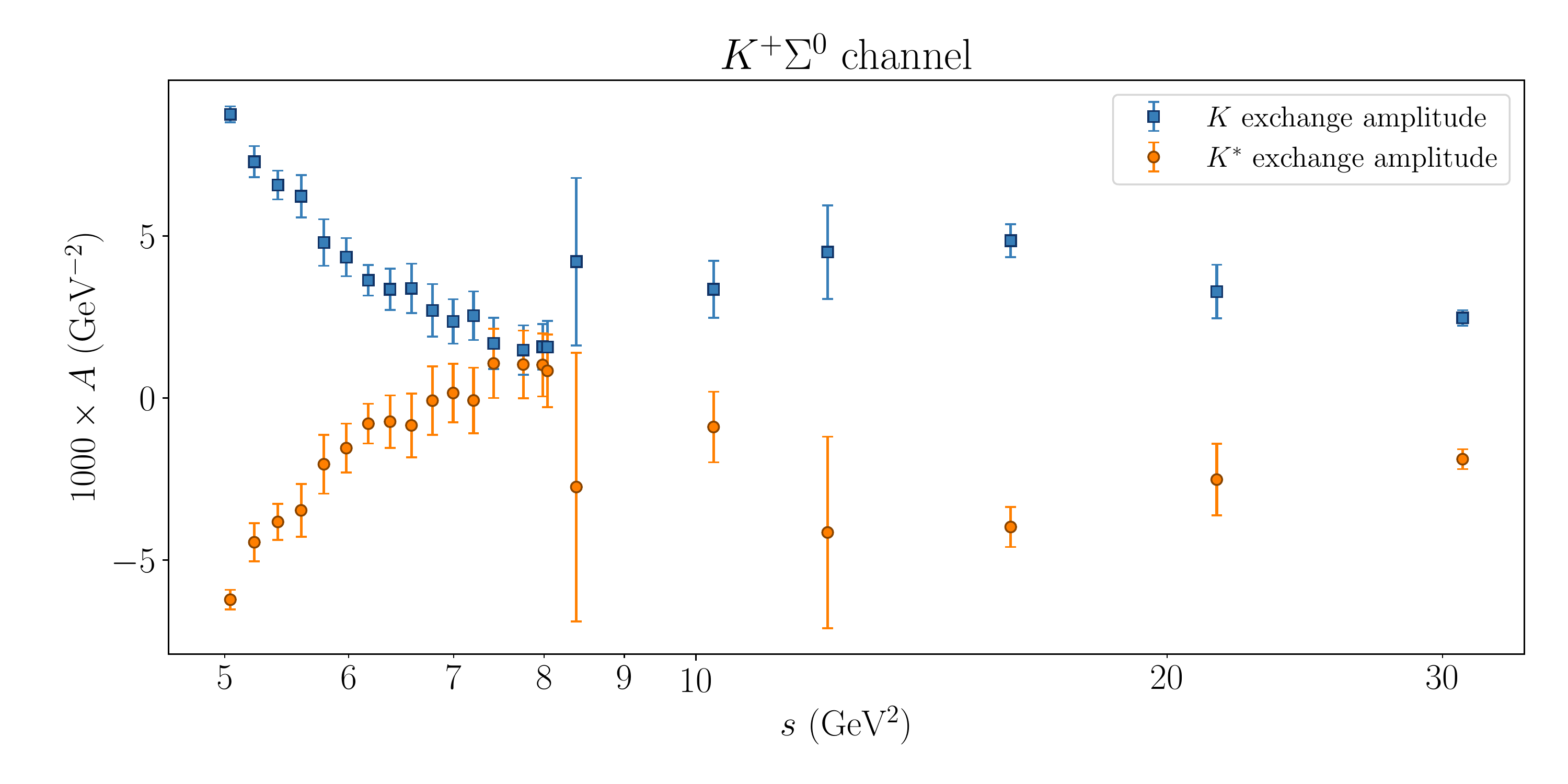}
  \end{subfigure}
  \caption{
    The amplitudes $A_K(s)$ and $A_{K^*}(s)$
    extracted from fits to Eq.~(\ref{eqn:double}),
    plotted against Mandelstam $s$.
    CLAS data~\cite{McCracken:2009ra,Dey:2010hh}
    are plotted for $s < 8.1$~GeV$^2$,
    and SLAC data~\cite{Boyarski:1970yc,Anderson:1976ph,Quinn:1979zp}
    for higher $s$.
    To improve readability,
    every fourth CLAS data point was plotted.
  }
  \label{fig:A1A2}
\end{figure*}

In Fig.~\ref{fig:A1A2}, we can see the results of
the double-exponential fit plotted against $s$.
In accordance with our previous hypothesis,
$K^+\Lambda$ production appears to occur
predominantly through $K$ Reggeon exchange,
in the $5 < s < 8.1$~GeV$^2$ region,
while $K^+\Sigma^0$ production requires
a mix of $K$ and $K^*$ Reggeon exchange at all $s$.
Moreover, we have found that the $K$ and $K^*$ Reggeon
exchange contributions are opposite in phase.
Additionally, at $s \approx 8$~GeV$^2$,
the exchange behavior of both channels does appear to change,
with $K^*$ exchange suddenly becoming significant
in $K^+\Lambda$ production.

One of the peculiarities in the bottom panel of Fig.~\ref{fig:A1A2}
is the apparent convergence of the amplitudes
for the $K^+\Sigma^0$ channel $s\approx8$~GeV$^2$.
We cannot say whether this is a robust conclusion of our analysis.
The simple form of Eq.~(\ref{eqn:double}),
or perhaps the other approximations used in this work,
may be inadequate for describing the $K^+\Sigma^0$ channel.
This channel did, after all, also evade description by the simple
Reggeon exchange model previously where the $K^+\Lambda$
proved amenable to such a description.

It is worth emphasizing that
the double-exponential fit performed here is somewhat crude,
and should not be taken as more than a preliminary investigation.
A theoretical analysis of the data from~\cite{Boyarski:1970yc}
was previously done in~\cite{Guidal:1997hy},
and the conclusions reached for these data were that $K$ and $K^*$ Reggeons
contributed about equally (with opposite phases) to the $K^+\Lambda$ channel,
and that the $K^*$ Reggeon contribution to $K^+\Sigma^0$ production
overwhelmed the $K$ contribution.
Our double-exponential fit is compatible with the previous observation
that $K$ and $K^*$ exchange both contribute significantly
to $K^+\Lambda$ production at large $s$,
but we do not observe the $K^*$ exchange amplitude overwhelming the
$K$ exchange amplitude for $K^+\Sigma^0$ production.
The conclusions of~\cite{Guidal:1997hy} regarding the $K^+\Sigma^0$
channel are likely more robust than ours,
but we take the consilience between~\cite{Guidal:1997hy}
and this work for the $K^+\Lambda$ channel
to lend credence to our analysis of this channel.


\section{Conclusions and Outlook}
\label{sec:end}

We have fit recent high-precision CLAS data for $K^+$ meson photoproduction
to the the exponential fit form $Ae^{Bt}$,
extracting the $t$-slope factor $B$ as a function of $s$
for the $K^+\Lambda$ and $K^+\Sigma^0$ channels.
In performing the extraction,
we developed a scheme for determining the range of $t$ to which the fit
should be applied,
and for estimating the systematic uncertainty due to our choice.

We found that $K^+\Lambda$ photoproduction is compatible with
exchange of a $K^+$ Reggeon being the dominant contribution
in the region $5 < s < 8.1$~GeV$^2$.
On the other hand, $K^+\Sigma^0$ production could not be attributed
to single-Reggeon exchange.
We then performed a fit to interfering exponential amplitudes
(using Eq.~(\ref{eqn:double})),
the results of which vindicate our conclusions from the single exponential fit.

At $s\approx 8$~GeV$^2$, which coincides with the limit of
recent high-precision CLAS data,
the behavior of the $t$-slope $B$ appears to change dramatically.
However, the world data at $s > 8$~GeV$^2$
is sparse and imprecise.
A better understanding whether and how
the production mechanisms for $K^+$ meson photoproduction
change at $s \approx 8$~GeV$^2$ will require obtaining more
high-precision data at and above this energy.
This will be possible at Jefferson Lab with the 12~GeV upgrade.

The analysis done in this work is
somewhat crude and strictly phenomenological.
A more complete analysis would account for $t$ dependence in the
differential cross section other than what is contained
in the functional form $Ae^{Bt}$,
including $t$-dependence of the parameter $s_0$ used in
the theoretical expression (\ref{eqn:B}) for $B$.
Additionally, a more complete analysis would account for
$s$-channel (resonance) contributions.

Nonetheless, our results hint at the production mechanisms
that may contribute to $K^+$ meson photoproduction at
intermediate energies.
Additionally, phenomenological $t$-slopes such as those we
have extracted may be potentially useful in a
Monte Carlo generator for $t$-channel production of hyperon channels.
For this reason, we have included a spreadsheet containing
the extracted $t$-slopes as supplemental material.

\section*{Acknowledgments}

This work was supported by DOE award {DE-SC0013620}.
R.B.\ was supported, in part,
by the FIU Graduate School through a Dissertation Year Fellowship.
A.F.\ was supported during the revision by ANL LDRD Project No.\ 2017-058-N0.
We would like to thank Jan Ryckebusch
for interesting discussions that helped contribute to this research,
and Seamus Riordan for reading the manuscript.

\bibliographystyle{apsrev4-1}
\bibliography{references}

\begin{thebibliography}{26}%
\makeatletter
\providecommand \@ifxundefined [1]{%
 \@ifx{#1\undefined}
}%
\providecommand \@ifnum [1]{%
 \ifnum #1\expandafter \@firstoftwo
 \else \expandafter \@secondoftwo
 \fi
}%
\providecommand \@ifx [1]{%
 \ifx #1\expandafter \@firstoftwo
 \else \expandafter \@secondoftwo
 \fi
}%
\providecommand \natexlab [1]{#1}%
\providecommand \enquote  [1]{``#1''}%
\providecommand \bibnamefont  [1]{#1}%
\providecommand \bibfnamefont [1]{#1}%
\providecommand \citenamefont [1]{#1}%
\providecommand \href@noop [0]{\@secondoftwo}%
\providecommand \href [0]{\begingroup \@sanitize@url \@href}%
\providecommand \@href[1]{\@@startlink{#1}\@@href}%
\providecommand \@@href[1]{\endgroup#1\@@endlink}%
\providecommand \@sanitize@url [0]{\catcode `\\12\catcode `\$12\catcode
  `\&12\catcode `\#12\catcode `\^12\catcode `\_12\catcode `\%12\relax}%
\providecommand \@@startlink[1]{}%
\providecommand \@@endlink[0]{}%
\providecommand \url  [0]{\begingroup\@sanitize@url \@url }%
\providecommand \@url [1]{\endgroup\@href {#1}{\urlprefix }}%
\providecommand \urlprefix  [0]{URL }%
\providecommand \Eprint [0]{\href }%
\providecommand \doibase [0]{http://dx.doi.org/}%
\providecommand \selectlanguage [0]{\@gobble}%
\providecommand \bibinfo  [0]{\@secondoftwo}%
\providecommand \bibfield  [0]{\@secondoftwo}%
\providecommand \translation [1]{[#1]}%
\providecommand \BibitemOpen [0]{}%
\providecommand \bibitemStop [0]{}%
\providecommand \bibitemNoStop [0]{.\EOS\space}%
\providecommand \EOS [0]{\spacefactor3000\relax}%
\providecommand \BibitemShut  [1]{\csname bibitem#1\endcsname}%
\let\auto@bib@innerbib\@empty
\bibitem [{\citenamefont {Bradford}\ \emph {et~al.}(2006)\citenamefont
  {Bradford} \emph {et~al.}}]{Bradford:2005pt}%
  \BibitemOpen
  \bibfield  {author} {\bibinfo {author} {\bibfnamefont {R.}~\bibnamefont
  {Bradford}} \emph {et~al.} (\bibinfo {collaboration} {CLAS}),\ }\href
  {\doibase 10.1103/PhysRevC.73.035202} {\bibfield  {journal} {\bibinfo
  {journal} {Phys. Rev.}\ }\textbf {\bibinfo {volume} {C73}},\ \bibinfo {pages}
  {035202} (\bibinfo {year} {2006})},\ \Eprint
  {http://arxiv.org/abs/nucl-ex/0509033} {arXiv:nucl-ex/0509033 [nucl-ex]}
  \BibitemShut {NoStop}%
\bibitem [{\citenamefont {McCracken}\ \emph {et~al.}(2010)\citenamefont
  {McCracken} \emph {et~al.}}]{McCracken:2009ra}%
  \BibitemOpen
  \bibfield  {author} {\bibinfo {author} {\bibfnamefont {M.~E.}\ \bibnamefont
  {McCracken}} \emph {et~al.} (\bibinfo {collaboration} {CLAS}),\ }\href
  {\doibase 10.1103/PhysRevC.81.025201} {\bibfield  {journal} {\bibinfo
  {journal} {Phys. Rev.}\ }\textbf {\bibinfo {volume} {C81}},\ \bibinfo {pages}
  {025201} (\bibinfo {year} {2010})},\ \Eprint {http://arxiv.org/abs/0912.4274}
  {arXiv:0912.4274 [nucl-ex]} \BibitemShut {NoStop}%
\bibitem [{\citenamefont {Dey}\ \emph {et~al.}(2010)\citenamefont {Dey} \emph
  {et~al.}}]{Dey:2010hh}%
  \BibitemOpen
  \bibfield  {author} {\bibinfo {author} {\bibfnamefont {B.}~\bibnamefont
  {Dey}} \emph {et~al.} (\bibinfo {collaboration} {CLAS}),\ }\href {\doibase
  10.1103/PhysRevC.82.025202} {\bibfield  {journal} {\bibinfo  {journal} {Phys.
  Rev.}\ }\textbf {\bibinfo {volume} {C82}},\ \bibinfo {pages} {025202}
  (\bibinfo {year} {2010})},\ \Eprint {http://arxiv.org/abs/1006.0374}
  {arXiv:1006.0374 [nucl-ex]} \BibitemShut {NoStop}%
\bibitem [{\citenamefont {Schumacher}\ and\ \citenamefont
  {Sargsian}(2011)}]{Schumacher:2010qx}%
  \BibitemOpen
  \bibfield  {author} {\bibinfo {author} {\bibfnamefont {R.~A.}\ \bibnamefont
  {Schumacher}}\ and\ \bibinfo {author} {\bibfnamefont {M.~M.}\ \bibnamefont
  {Sargsian}},\ }\href {\doibase 10.1103/PhysRevC.83.025207} {\bibfield
  {journal} {\bibinfo  {journal} {Phys. Rev.}\ }\textbf {\bibinfo {volume}
  {C83}},\ \bibinfo {pages} {025207} (\bibinfo {year} {2011})},\ \Eprint
  {http://arxiv.org/abs/1012.2126} {arXiv:1012.2126 [hep-ph]} \BibitemShut
  {NoStop}%
\bibitem [{\citenamefont {Anisovich}\ \emph {et~al.}(2007)\citenamefont
  {Anisovich}, \citenamefont {Kleber}, \citenamefont {Klempt}, \citenamefont
  {Nikonov}, \citenamefont {Sarantsev},\ and\ \citenamefont
  {Thoma}}]{Anisovich:2007bq}%
  \BibitemOpen
  \bibfield  {author} {\bibinfo {author} {\bibfnamefont {A.~V.}\ \bibnamefont
  {Anisovich}}, \bibinfo {author} {\bibfnamefont {V.}~\bibnamefont {Kleber}},
  \bibinfo {author} {\bibfnamefont {E.}~\bibnamefont {Klempt}}, \bibinfo
  {author} {\bibfnamefont {V.~A.}\ \bibnamefont {Nikonov}}, \bibinfo {author}
  {\bibfnamefont {A.~V.}\ \bibnamefont {Sarantsev}}, \ and\ \bibinfo {author}
  {\bibfnamefont {U.}~\bibnamefont {Thoma}},\ }\href {\doibase
  10.1140/epja/i2007-10503-6} {\bibfield  {journal} {\bibinfo  {journal} {Eur.
  Phys. J.}\ }\textbf {\bibinfo {volume} {A34}},\ \bibinfo {pages} {243}
  (\bibinfo {year} {2007})},\ \Eprint {http://arxiv.org/abs/0707.3596}
  {arXiv:0707.3596 [hep-ph]} \BibitemShut {NoStop}%
\bibitem [{\citenamefont {Nikonov}\ \emph {et~al.}(2008)\citenamefont
  {Nikonov}, \citenamefont {Anisovich}, \citenamefont {Klempt}, \citenamefont
  {Sarantsev},\ and\ \citenamefont {Thoma}}]{Nikonov:2007br}%
  \BibitemOpen
  \bibfield  {author} {\bibinfo {author} {\bibfnamefont {V.~A.}\ \bibnamefont
  {Nikonov}}, \bibinfo {author} {\bibfnamefont {A.~V.}\ \bibnamefont
  {Anisovich}}, \bibinfo {author} {\bibfnamefont {E.}~\bibnamefont {Klempt}},
  \bibinfo {author} {\bibfnamefont {A.~V.}\ \bibnamefont {Sarantsev}}, \ and\
  \bibinfo {author} {\bibfnamefont {U.}~\bibnamefont {Thoma}},\ }\href
  {\doibase 10.1016/j.physletb.2008.03.004} {\bibfield  {journal} {\bibinfo
  {journal} {Phys. Lett.}\ }\textbf {\bibinfo {volume} {B662}},\ \bibinfo
  {pages} {245} (\bibinfo {year} {2008})},\ \Eprint
  {http://arxiv.org/abs/0707.3600} {arXiv:0707.3600 [hep-ph]} \BibitemShut
  {NoStop}%
\bibitem [{\citenamefont {Anisovich}\ \emph
  {et~al.}(2011{\natexlab{a}})\citenamefont {Anisovich}, \citenamefont
  {Klempt}, \citenamefont {Nikonov}, \citenamefont {Sarantsev},\ and\
  \citenamefont {Thoma}}]{Anisovich:2010an}%
  \BibitemOpen
  \bibfield  {author} {\bibinfo {author} {\bibfnamefont {A.~V.}\ \bibnamefont
  {Anisovich}}, \bibinfo {author} {\bibfnamefont {E.}~\bibnamefont {Klempt}},
  \bibinfo {author} {\bibfnamefont {V.~A.}\ \bibnamefont {Nikonov}}, \bibinfo
  {author} {\bibfnamefont {A.~V.}\ \bibnamefont {Sarantsev}}, \ and\ \bibinfo
  {author} {\bibfnamefont {U.}~\bibnamefont {Thoma}},\ }\href {\doibase
  10.1140/epja/i2011-11027-2} {\bibfield  {journal} {\bibinfo  {journal} {Eur.
  Phys. J.}\ }\textbf {\bibinfo {volume} {A47}},\ \bibinfo {pages} {27}
  (\bibinfo {year} {2011}{\natexlab{a}})},\ \Eprint
  {http://arxiv.org/abs/1009.4803} {arXiv:1009.4803 [hep-ph]} \BibitemShut
  {NoStop}%
\bibitem [{\citenamefont {Anisovich}\ \emph
  {et~al.}(2011{\natexlab{b}})\citenamefont {Anisovich}, \citenamefont
  {Klempt}, \citenamefont {Nikonov}, \citenamefont {Sarantsev},\ and\
  \citenamefont {Thoma}}]{Anisovich:2011ye}%
  \BibitemOpen
  \bibfield  {author} {\bibinfo {author} {\bibfnamefont {A.~V.}\ \bibnamefont
  {Anisovich}}, \bibinfo {author} {\bibfnamefont {E.}~\bibnamefont {Klempt}},
  \bibinfo {author} {\bibfnamefont {V.~A.}\ \bibnamefont {Nikonov}}, \bibinfo
  {author} {\bibfnamefont {A.~V.}\ \bibnamefont {Sarantsev}}, \ and\ \bibinfo
  {author} {\bibfnamefont {U.}~\bibnamefont {Thoma}},\ }\href {\doibase
  10.1140/epja/i2011-11153-9} {\bibfield  {journal} {\bibinfo  {journal} {Eur.
  Phys. J.}\ }\textbf {\bibinfo {volume} {A47}},\ \bibinfo {pages} {153}
  (\bibinfo {year} {2011}{\natexlab{b}})},\ \Eprint
  {http://arxiv.org/abs/1109.0970} {arXiv:1109.0970 [hep-ph]} \BibitemShut
  {NoStop}%
\bibitem [{\citenamefont {Anisovich}\ \emph {et~al.}(2012)\citenamefont
  {Anisovich}, \citenamefont {Klempt}, \citenamefont {Nikonov}, \citenamefont
  {Sarantsev},\ and\ \citenamefont {Thoma}}]{Anisovich:2011su}%
  \BibitemOpen
  \bibfield  {author} {\bibinfo {author} {\bibfnamefont {A.~V.}\ \bibnamefont
  {Anisovich}}, \bibinfo {author} {\bibfnamefont {E.}~\bibnamefont {Klempt}},
  \bibinfo {author} {\bibfnamefont {V.~A.}\ \bibnamefont {Nikonov}}, \bibinfo
  {author} {\bibfnamefont {A.~V.}\ \bibnamefont {Sarantsev}}, \ and\ \bibinfo
  {author} {\bibfnamefont {U.}~\bibnamefont {Thoma}},\ }\href {\doibase
  10.1016/j.physletb.2012.03.066} {\bibfield  {journal} {\bibinfo  {journal}
  {Phys. Lett.}\ }\textbf {\bibinfo {volume} {B711}},\ \bibinfo {pages} {167}
  (\bibinfo {year} {2012})},\ \Eprint {http://arxiv.org/abs/1111.6150}
  {arXiv:1111.6150 [hep-ex]} \BibitemShut {NoStop}%
\bibitem [{\citenamefont {Anisovich}\ \emph {et~al.}(2014)\citenamefont
  {Anisovich}, \citenamefont {Beck}, \citenamefont {Burkert}, \citenamefont
  {Klempt}, \citenamefont {McCracken}, \citenamefont {Nikonov}, \citenamefont
  {Sarantsev}, \citenamefont {Schumacher},\ and\ \citenamefont
  {Thoma}}]{Anisovich:2014yza}%
  \BibitemOpen
  \bibfield  {author} {\bibinfo {author} {\bibfnamefont {A.~V.}\ \bibnamefont
  {Anisovich}}, \bibinfo {author} {\bibfnamefont {R.}~\bibnamefont {Beck}},
  \bibinfo {author} {\bibfnamefont {V.}~\bibnamefont {Burkert}}, \bibinfo
  {author} {\bibfnamefont {E.}~\bibnamefont {Klempt}}, \bibinfo {author}
  {\bibfnamefont {M.~E.}\ \bibnamefont {McCracken}}, \bibinfo {author}
  {\bibfnamefont {V.~A.}\ \bibnamefont {Nikonov}}, \bibinfo {author}
  {\bibfnamefont {A.~V.}\ \bibnamefont {Sarantsev}}, \bibinfo {author}
  {\bibfnamefont {R.~A.}\ \bibnamefont {Schumacher}}, \ and\ \bibinfo {author}
  {\bibfnamefont {U.}~\bibnamefont {Thoma}},\ }\href {\doibase
  10.1140/epja/i2014-14129-3} {\bibfield  {journal} {\bibinfo  {journal} {Eur.
  Phys. J.}\ }\textbf {\bibinfo {volume} {A50}},\ \bibinfo {pages} {129}
  (\bibinfo {year} {2014})},\ \Eprint {http://arxiv.org/abs/1404.4587}
  {arXiv:1404.4587 [nucl-ex]} \BibitemShut {NoStop}%
\bibitem [{\citenamefont {Bauer}\ \emph {et~al.}(1978)\citenamefont {Bauer},
  \citenamefont {Spital}, \citenamefont {Yennie},\ and\ \citenamefont
  {Pipkin}}]{Bauer:1977iq}%
  \BibitemOpen
  \bibfield  {author} {\bibinfo {author} {\bibfnamefont {T.~H.}\ \bibnamefont
  {Bauer}}, \bibinfo {author} {\bibfnamefont {R.~D.}\ \bibnamefont {Spital}},
  \bibinfo {author} {\bibfnamefont {D.~R.}\ \bibnamefont {Yennie}}, \ and\
  \bibinfo {author} {\bibfnamefont {F.~M.}\ \bibnamefont {Pipkin}},\ }\href
  {\doibase 10.1103/RevModPhys.50.261} {\bibfield  {journal} {\bibinfo
  {journal} {Rev. Mod. Phys.}\ }\textbf {\bibinfo {volume} {50}},\ \bibinfo
  {pages} {261} (\bibinfo {year} {1978})}\BibitemShut {NoStop}%
\bibitem [{\citenamefont {Collins}(2009)}]{Collins:1977jy}%
  \BibitemOpen
  \bibfield  {author} {\bibinfo {author} {\bibfnamefont {P.~D.~B.}\
  \bibnamefont {Collins}},\ }\href
  {http://www-spires.fnal.gov/spires/find/books/www?cl=QC793.3.R4C695} {\emph
  {\bibinfo {title} {{An Introduction to Regge Theory and High-Energy
  Physics}}}},\ Cambridge Monographs on Mathematical Physics\ (\bibinfo
  {publisher} {Cambridge Univ. Press},\ \bibinfo {address} {Cambridge, UK},\
  \bibinfo {year} {2009})\BibitemShut {NoStop}%
\bibitem [{\citenamefont {Guidal}\ \emph {et~al.}(1997)\citenamefont {Guidal},
  \citenamefont {Laget},\ and\ \citenamefont {Vanderhaeghen}}]{Guidal:1997hy}%
  \BibitemOpen
  \bibfield  {author} {\bibinfo {author} {\bibfnamefont {M.}~\bibnamefont
  {Guidal}}, \bibinfo {author} {\bibfnamefont {J.~M.}\ \bibnamefont {Laget}}, \
  and\ \bibinfo {author} {\bibfnamefont {M.}~\bibnamefont {Vanderhaeghen}},\
  }\href {\doibase 10.1016/S0375-9474(97)00612-X} {\bibfield  {journal}
  {\bibinfo  {journal} {Nucl. Phys.}\ }\textbf {\bibinfo {volume} {A627}},\
  \bibinfo {pages} {645} (\bibinfo {year} {1997})}\BibitemShut {NoStop}%
\bibitem [{\citenamefont {Boyarski}\ \emph {et~al.}(1971)\citenamefont
  {Boyarski}, \citenamefont {Diebold}, \citenamefont {Ecklund}, \citenamefont
  {Fischer}, \citenamefont {Murata}, \citenamefont {Richter},\ and\
  \citenamefont {Sands}}]{Boyarski:1970yc}%
  \BibitemOpen
  \bibfield  {author} {\bibinfo {author} {\bibfnamefont {A.}~\bibnamefont
  {Boyarski}}, \bibinfo {author} {\bibfnamefont {R.~E.}\ \bibnamefont
  {Diebold}}, \bibinfo {author} {\bibfnamefont {S.~D.}\ \bibnamefont
  {Ecklund}}, \bibinfo {author} {\bibfnamefont {G.~E.}\ \bibnamefont
  {Fischer}}, \bibinfo {author} {\bibfnamefont {Y.}~\bibnamefont {Murata}},
  \bibinfo {author} {\bibfnamefont {B.}~\bibnamefont {Richter}}, \ and\
  \bibinfo {author} {\bibfnamefont {M.}~\bibnamefont {Sands}},\ }\href
  {\doibase 10.1016/0370-2693(71)90677-0} {\bibfield  {journal} {\bibinfo
  {journal} {Phys. Lett.}\ }\textbf {\bibinfo {volume} {B34}},\ \bibinfo
  {pages} {547} (\bibinfo {year} {1971})}\BibitemShut {NoStop}%
\bibitem [{\citenamefont {Behrend}\ \emph {et~al.}(1978)\citenamefont
  {Behrend}, \citenamefont {Bodenkamp}, \citenamefont {Hesse}, \citenamefont
  {McNeely}, \citenamefont {Miyachi}, \citenamefont {Fries}, \citenamefont
  {Heine}, \citenamefont {Hirschmann}, \citenamefont {Markou},\ and\
  \citenamefont {Seitz}}]{Behrend:1978ik}%
  \BibitemOpen
  \bibfield  {author} {\bibinfo {author} {\bibfnamefont {H.~J.}\ \bibnamefont
  {Behrend}}, \bibinfo {author} {\bibfnamefont {J.}~\bibnamefont {Bodenkamp}},
  \bibinfo {author} {\bibfnamefont {W.~P.}\ \bibnamefont {Hesse}}, \bibinfo
  {author} {\bibfnamefont {W.~A.}\ \bibnamefont {McNeely}, \bibfnamefont
  {Jr.}}, \bibinfo {author} {\bibfnamefont {T.}~\bibnamefont {Miyachi}},
  \bibinfo {author} {\bibfnamefont {D.~C.}\ \bibnamefont {Fries}}, \bibinfo
  {author} {\bibfnamefont {P.}~\bibnamefont {Heine}}, \bibinfo {author}
  {\bibfnamefont {H.}~\bibnamefont {Hirschmann}}, \bibinfo {author}
  {\bibfnamefont {A.}~\bibnamefont {Markou}}, \ and\ \bibinfo {author}
  {\bibfnamefont {E.}~\bibnamefont {Seitz}},\ }\href {\doibase
  10.1016/0550-3213(78)90497-2} {\bibfield  {journal} {\bibinfo  {journal}
  {Nucl. Phys.}\ }\textbf {\bibinfo {volume} {B144}},\ \bibinfo {pages} {22}
  (\bibinfo {year} {1978})}\BibitemShut {NoStop}%
\bibitem [{\citenamefont {Seraydaryan}\ \emph {et~al.}(2014)\citenamefont
  {Seraydaryan} \emph {et~al.}}]{Seraydaryan:2013ija}%
  \BibitemOpen
  \bibfield  {author} {\bibinfo {author} {\bibfnamefont {H.}~\bibnamefont
  {Seraydaryan}} \emph {et~al.} (\bibinfo {collaboration} {CLAS}),\ }\href
  {\doibase 10.1103/PhysRevC.89.055206} {\bibfield  {journal} {\bibinfo
  {journal} {Phys. Rev.}\ }\textbf {\bibinfo {volume} {C89}},\ \bibinfo {pages}
  {055206} (\bibinfo {year} {2014})},\ \Eprint {http://arxiv.org/abs/1308.1363}
  {arXiv:1308.1363 [hep-ex]} \BibitemShut {NoStop}%
\bibitem [{\citenamefont {Dey}\ \emph {et~al.}(2014)\citenamefont {Dey},
  \citenamefont {Meyer}, \citenamefont {Bellis},\ and\ \citenamefont
  {Williams}}]{Dey:2014tfa}%
  \BibitemOpen
  \bibfield  {author} {\bibinfo {author} {\bibfnamefont {B.}~\bibnamefont
  {Dey}}, \bibinfo {author} {\bibfnamefont {C.~A.}\ \bibnamefont {Meyer}},
  \bibinfo {author} {\bibfnamefont {M.}~\bibnamefont {Bellis}}, \ and\ \bibinfo
  {author} {\bibfnamefont {M.}~\bibnamefont {Williams}} (\bibinfo
  {collaboration} {CLAS}),\ }\href {\doibase 10.1103/PhysRevC.90.019901,
  10.1103/PhysRevC.89.055208} {\bibfield  {journal} {\bibinfo  {journal} {Phys.
  Rev.}\ }\textbf {\bibinfo {volume} {C89}},\ \bibinfo {pages} {055208}
  (\bibinfo {year} {2014})},\ \bibinfo {note} {[Addendum: Phys.
  Rev.C90,no.1,019901(2014)]},\ \Eprint {http://arxiv.org/abs/1403.2110}
  {arXiv:1403.2110 [nucl-ex]} \BibitemShut {NoStop}%
\bibitem [{\citenamefont {Elings}\ \emph {et~al.}(1967)\citenamefont {Elings},
  \citenamefont {Cohen}, \citenamefont {Garelick}, \citenamefont {Homma},
  \citenamefont {Lewis}, \citenamefont {Lobar}, \citenamefont {Luckey},\ and\
  \citenamefont {Osborne}}]{Elings:1967af}%
  \BibitemOpen
  \bibfield  {author} {\bibinfo {author} {\bibfnamefont {V.~B.}\ \bibnamefont
  {Elings}}, \bibinfo {author} {\bibfnamefont {K.~J.}\ \bibnamefont {Cohen}},
  \bibinfo {author} {\bibfnamefont {D.~A.}\ \bibnamefont {Garelick}}, \bibinfo
  {author} {\bibfnamefont {S.}~\bibnamefont {Homma}}, \bibinfo {author}
  {\bibfnamefont {R.~A.}\ \bibnamefont {Lewis}}, \bibinfo {author}
  {\bibfnamefont {W.}~\bibnamefont {Lobar}}, \bibinfo {author} {\bibfnamefont
  {D.}~\bibnamefont {Luckey}}, \ and\ \bibinfo {author} {\bibfnamefont {L.~S.}\
  \bibnamefont {Osborne}},\ }\href {\doibase 10.1103/PhysRev.156.1433}
  {\bibfield  {journal} {\bibinfo  {journal} {Phys. Rev.}\ }\textbf {\bibinfo
  {volume} {156}},\ \bibinfo {pages} {1433} (\bibinfo {year}
  {1967})}\BibitemShut {NoStop}%
\bibitem [{\citenamefont {Bleckmann}\ \emph {et~al.}(1970)\citenamefont
  {Bleckmann}, \citenamefont {Herda}, \citenamefont {Opara}, \citenamefont
  {Schulz}, \citenamefont {Schwille},\ and\ \citenamefont
  {Urbahn}}]{Bleckmann:1970kb}%
  \BibitemOpen
  \bibfield  {author} {\bibinfo {author} {\bibfnamefont {A.}~\bibnamefont
  {Bleckmann}}, \bibinfo {author} {\bibfnamefont {S.}~\bibnamefont {Herda}},
  \bibinfo {author} {\bibfnamefont {U.}~\bibnamefont {Opara}}, \bibinfo
  {author} {\bibfnamefont {W.}~\bibnamefont {Schulz}}, \bibinfo {author}
  {\bibfnamefont {W.~J.}\ \bibnamefont {Schwille}}, \ and\ \bibinfo {author}
  {\bibfnamefont {H.}~\bibnamefont {Urbahn}},\ }\href {\doibase
  10.1007/BF01408507} {\bibfield  {journal} {\bibinfo  {journal} {Z. Phys.}\
  }\textbf {\bibinfo {volume} {239}},\ \bibinfo {pages} {1} (\bibinfo {year}
  {1970})}\BibitemShut {NoStop}%
\bibitem [{\citenamefont {Anderson}\ \emph {et~al.}(1976)\citenamefont
  {Anderson}, \citenamefont {Gustavson}, \citenamefont {Ritson}, \citenamefont
  {Weitsch}, \citenamefont {Halpern}, \citenamefont {Prepost}, \citenamefont
  {Tompkins},\ and\ \citenamefont {Wiser}}]{Anderson:1976ph}%
  \BibitemOpen
  \bibfield  {author} {\bibinfo {author} {\bibfnamefont {R.~L.}\ \bibnamefont
  {Anderson}}, \bibinfo {author} {\bibfnamefont {D.}~\bibnamefont {Gustavson}},
  \bibinfo {author} {\bibfnamefont {D.}~\bibnamefont {Ritson}}, \bibinfo
  {author} {\bibfnamefont {G.~A.}\ \bibnamefont {Weitsch}}, \bibinfo {author}
  {\bibfnamefont {H.~J.}\ \bibnamefont {Halpern}}, \bibinfo {author}
  {\bibfnamefont {R.}~\bibnamefont {Prepost}}, \bibinfo {author} {\bibfnamefont
  {D.~H.}\ \bibnamefont {Tompkins}}, \ and\ \bibinfo {author} {\bibfnamefont
  {D.~E.}\ \bibnamefont {Wiser}},\ }\href {\doibase 10.1103/PhysRevD.14.679}
  {\bibfield  {journal} {\bibinfo  {journal} {Phys. Rev.}\ }\textbf {\bibinfo
  {volume} {D14}},\ \bibinfo {pages} {679} (\bibinfo {year}
  {1976})}\BibitemShut {NoStop}%
\bibitem [{\citenamefont {Quinn}\ \emph {et~al.}(1979)\citenamefont {Quinn},
  \citenamefont {Rutherfoord}, \citenamefont {Shupe}, \citenamefont {Sherden},
  \citenamefont {Siemann},\ and\ \citenamefont {Sinclair}}]{Quinn:1979zp}%
  \BibitemOpen
  \bibfield  {author} {\bibinfo {author} {\bibfnamefont {D.~J.}\ \bibnamefont
  {Quinn}}, \bibinfo {author} {\bibfnamefont {J.~P.}\ \bibnamefont
  {Rutherfoord}}, \bibinfo {author} {\bibfnamefont {M.~A.}\ \bibnamefont
  {Shupe}}, \bibinfo {author} {\bibfnamefont {D.}~\bibnamefont {Sherden}},
  \bibinfo {author} {\bibfnamefont {R.}~\bibnamefont {Siemann}}, \ and\
  \bibinfo {author} {\bibfnamefont {C.~K.}\ \bibnamefont {Sinclair}},\ }\href
  {\doibase 10.1103/PhysRevD.20.1553} {\bibfield  {journal} {\bibinfo
  {journal} {Phys. Rev.}\ }\textbf {\bibinfo {volume} {D20}},\ \bibinfo {pages}
  {1553} (\bibinfo {year} {1979})}\BibitemShut {NoStop}%
\bibitem [{\citenamefont {Bockhorst}\ \emph {et~al.}(1994)\citenamefont
  {Bockhorst} \emph {et~al.}}]{Bockhorst:1994jf}%
  \BibitemOpen
  \bibfield  {author} {\bibinfo {author} {\bibfnamefont {M.}~\bibnamefont
  {Bockhorst}} \emph {et~al.},\ }\href {\doibase 10.1007/BF01577542} {\bibfield
   {journal} {\bibinfo  {journal} {Z. Phys.}\ }\textbf {\bibinfo {volume}
  {C63}},\ \bibinfo {pages} {37} (\bibinfo {year} {1994})}\BibitemShut
  {NoStop}%
\bibitem [{\citenamefont {Tran}\ \emph {et~al.}(1998)\citenamefont {Tran} \emph
  {et~al.}}]{Tran:1998qw}%
  \BibitemOpen
  \bibfield  {author} {\bibinfo {author} {\bibfnamefont {M.~Q.}\ \bibnamefont
  {Tran}} \emph {et~al.} (\bibinfo {collaboration} {SAPHIR}),\ }\href {\doibase
  10.1016/S0370-2693(98)01393-8} {\bibfield  {journal} {\bibinfo  {journal}
  {Phys. Lett.}\ }\textbf {\bibinfo {volume} {B445}},\ \bibinfo {pages} {20}
  (\bibinfo {year} {1998})}\BibitemShut {NoStop}%
\bibitem [{\citenamefont {Glander}\ \emph {et~al.}(2004)\citenamefont {Glander}
  \emph {et~al.}}]{Glander:2003jw}%
  \BibitemOpen
  \bibfield  {author} {\bibinfo {author} {\bibfnamefont {K.~H.}\ \bibnamefont
  {Glander}} \emph {et~al.},\ }\href {\doibase 10.1140/epja/i2003-10119-x}
  {\bibfield  {journal} {\bibinfo  {journal} {Eur. Phys. J.}\ }\textbf
  {\bibinfo {volume} {A19}},\ \bibinfo {pages} {251} (\bibinfo {year}
  {2004})},\ \Eprint {http://arxiv.org/abs/nucl-ex/0308025}
  {arXiv:nucl-ex/0308025 [nucl-ex]} \BibitemShut {NoStop}%
\bibitem [{\citenamefont {Corthals}\ \emph {et~al.}(2006)\citenamefont
  {Corthals}, \citenamefont {Ryckebusch},\ and\ \citenamefont
  {Van~Cauteren}}]{Corthals:2005ce}%
  \BibitemOpen
  \bibfield  {author} {\bibinfo {author} {\bibfnamefont {T.}~\bibnamefont
  {Corthals}}, \bibinfo {author} {\bibfnamefont {J.}~\bibnamefont
  {Ryckebusch}}, \ and\ \bibinfo {author} {\bibfnamefont {T.}~\bibnamefont
  {Van~Cauteren}},\ }\href {\doibase 10.1103/PhysRevC.73.045207} {\bibfield
  {journal} {\bibinfo  {journal} {Phys. Rev.}\ }\textbf {\bibinfo {volume}
  {C73}},\ \bibinfo {pages} {045207} (\bibinfo {year} {2006})},\ \Eprint
  {http://arxiv.org/abs/nucl-th/0510056} {arXiv:nucl-th/0510056 [nucl-th]}
  \BibitemShut {NoStop}%
\bibitem [{\citenamefont {Corthals}\ \emph {et~al.}(2007)\citenamefont
  {Corthals}, \citenamefont {Ireland}, \citenamefont {Van~Cauteren},\ and\
  \citenamefont {Ryckebusch}}]{Corthals:2006nz}%
  \BibitemOpen
  \bibfield  {author} {\bibinfo {author} {\bibfnamefont {T.}~\bibnamefont
  {Corthals}}, \bibinfo {author} {\bibfnamefont {D.~G.}\ \bibnamefont
  {Ireland}}, \bibinfo {author} {\bibfnamefont {T.}~\bibnamefont
  {Van~Cauteren}}, \ and\ \bibinfo {author} {\bibfnamefont {J.}~\bibnamefont
  {Ryckebusch}},\ }\href {\doibase 10.1103/PhysRevC.75.045204} {\bibfield
  {journal} {\bibinfo  {journal} {Phys. Rev.}\ }\textbf {\bibinfo {volume}
  {C75}},\ \bibinfo {pages} {045204} (\bibinfo {year} {2007})},\ \Eprint
  {http://arxiv.org/abs/nucl-th/0612085} {arXiv:nucl-th/0612085 [nucl-th]}
  \BibitemShut {NoStop}%
\end{thebibliography}%

\end{document}